\documentclass[%
reprint,
amsmath,amssymb,
prb,
]{revtex4-2}

\setcitestyle{numbers,square}
\usepackage{mathrsfs}
\usepackage{comment}

\usepackage{graphicx}% Include figure files
\usepackage{dcolumn}% Align table columns on decimal point
\usepackage{bm}% bold math
\usepackage{blindtext}

\usepackage{hyperref}
\usepackage{hypcap}
\hypersetup{colorlinks=true,citecolor=blue,linkcolor=blue,urlcolor=blue,filecolor=blue}

\newcommand{\comentar}[1]{}
\newcommand{\vb}[1] {\mathbf{#1}}

\newcommand{\scr}[1] {\mathcal{#1}}
\newcommand{\hb}[1] {\hat{\vb{#1}}}

\newcommand{\ten}[1] {\overleftrightarrow{\vb{#1}}}

\usepackage[dvipsnames]{xcolor} %for more colors

\newcommand{\rmi}[0]{{\rm i}}
\newcommand{\rme}[0]{{\rm e}}

\graphicspath{{Images/}}

\allowdisplaybreaks
\begin{document}
%--------------------------------------------%

%%%%%%%%%%%%%%%%%%%%%% TITLE AND ABSTRACT %%%%%%%%%%%%%%%%%%%%%%%%%
\title{Spectral density of angular momentum transfer from a swift electron to a large spherical nanoparticle}

%%%%%%%%%%%%%%%%%%%%%%%%%%%%%%%%%%%%%%%%%%%%%%%%%%%%%%%%%%%%%%%%%%%%%%%%%%%
\author{J. L. Briseño-Gómez$^1$}
\email{jorgeluisbrisenio@ciencias.unam.mx}
\author{A. Reyes-Coronado$^1$}
\affiliation{$^1$Departamento de F\'isica, Facultad de Ciencias, Universidad Nacional Aut\'onoma de M\'exico, Ciudad Universitaria, Av. Universidad $\# 3000$, Mexico City, 04510, Mexico.}
%%%%%%%%%%%%%%%%%%%%%%%%%%%%%%%%%%%%%%%%%%%%%%%%%%%%%%%%%%%%%%%%%%%%%%%%%%%
\date{\today}
%%%%%%%%%%%%%%%%%%%%%%%%%%%%%%%%%%%%%%%%%%%%%%%%%%%%%%%%%%%%%%%%%%%%%%%%%%%
\begin{abstract}
Swift electrons in scanning transmission electron microscopy transfer both linear and angular momentum to nanoparticles, underlying electron-beam-driven nanoscale manipulation (``electron tweezers''). Theoretical treatments of angular momentum transfer have so far relied either on the small-particle (dipolar) approximation, valid only well below the sizes typically manipulated experimentally, or on frequency-integrated multipolar calculations that leave the spectral structure of the interaction unresolved. Here we present a fully retarded, causal, multipole-converged electrodynamical methodology for the angular momentum transfer from a swift electron to an isolated spherical nanoparticle, based on a closed-surface Maxwell stress tensor formulation whose angular integrals reduce analytically to a small, material- and trajectory-independent set of irreducible integrals over associated Legendre functions. Exploiting the azimuthal selection rules of this decomposition lowers the cost of the double multipolar sum from $O(\ell_{\max}^4)$ to $O(\ell_{\max}^3)$, enabling convergence up to $\ell_{\max}=51$ for nanoparticles with radius as large as $a=50$~nm, nearly four times the multipole order reached in the largest previous calculation at this size and previously unreached for an optically complex, interband-dominated material, all at a wall-clock computational cost three to four orders of magnitude lower than that earlier, lower-order calculation. Applying this method to aluminum and gold nanoparticles up to $a=50$~nm, we resolve the spectral density of the angular momentum transfer across the full frequency domain and find that the angular momentum transfer is set by interference between the electron field and the field scattered by the nanoparticle, dominating the scattered-scattered contribution at essentially every frequency; the electric contribution dominates at moderate speeds, but the magnetic contribution, of the same sign, grows steadily in relative weight with electron speed, from a few percent of the total at $v=0.5c$ to between a quarter and a half of it at $v=0.95c$ when the electron trajectory passes close to the nanoparticle surface. At $a=50$~nm, gold transfers substantially more angular momentum than aluminum despite its markedly more intricate, interband-dominated response, by a factor that itself grows with electron speed, from about $2\times$ at $v=0.5c$ to about $5\times$ at $v=0.95c$ (at fixed impact parameter $b=51$~nm measured from the center of the nanoparticle).
\end{abstract}
%%%%%%%%%%%%%%%%%%%%%%%%%%%%%%%%%%%%%%%%%%%%%%%%%%%%%%%%%%%%%%%%%%%%
\maketitle

%%%%%%%%%%%%%%%%%%%%%%%%%%%%%%%%%%%%%%%%%%%%%%%%%%%%%%%%%%%%%%%%%%%%%%%%%%%
\section{Introduction}
%%%%%%%%%%%%%%%%%%%%%%%%%%%%%%%%%%%%%%%%%%%%%%%%%%%%%%%%%%%%%%%%%%%%%%%%%%%

Electron microscopy has become a central platform for interrogating matter at the nanoscale, with aberration-corrected imaging, electron energy-loss spectroscopy, and cathodoluminescence granting access to structural, electronic, and optical information otherwise unreachable at this length scale \cite{Batson0,GarciadeAbajo-1,garcia2021optical}.

Swift electrons can also transfer both linear and angular momentum to nanoparticles, harnessed for nanoscale mechanical manipulation (``electron tweezers") \cite{Oleshko,verbeeck2013,Batson01,GarciadeAbajo0,PRBCoronado}. As established in Ref.~\cite{castrejon2026electrodynamics} and summarized there in detail, accurate predictions require enforcing both causality of the dielectric response and convergence of the multipolar expansion \cite{Lagos2,castrejon2021time,castrejon2021effects}: within the small-particle limit, doing so yields a net transverse linear momentum transfer that is attractive toward the electron trajectory, in contrast to earlier claims of net repulsion \cite{castrejon2021effects}.

The extension of the method to study linear momentum transfer to large spherical NPs, where a fully retarded, multipolar description is indispensable, was carried out in Ref.~\cite{castrejon2026electrodynamics}. There, analytical expressions for the spectral density of linear momentum transfer were derived using a closed-surface Maxwell stress tensor formulation, and the angular integrals were reduced to a set of irreducible integrals over products of associated Legendre functions that are independent of material and geometric parameters. This structure enabled numerically efficient, machine-precision evaluation for NPs with radius as large as $a = 50$~nm, with multipole orders up to $\ell_{\max} = 50$, and confirmed that the net transverse linear momentum transfer remains attractive throughout the physically relevant parameter space within the isolated local spherical model.

A parallel and, so far, separate line of research has addressed the angular momentum transfer (AMT). Building on a general classical-electrodynamics formulation of the torque exerted by a swift electron on a nanoparticle \cite{castellanos2021angular}, Ref.~\cite{castellanos2023theory} derived the multipolar expression for the AMT to a spherical NP and used it to validate the small-particle (dipolar) approximation for aluminum, gold, and bismuth nanoparticles of radius $a=1$~nm, finding it accurate for impact parameters $b\gtrsim4a$ and electron speeds $v\gtrsim0.5c$. Pushing beyond the dipolar regime, that work further showed that the transferred angular momentum retains a constant sign throughout the parameter range explored and remains dominated by its electric contribution, extending the multipolar calculation to $a=50$~nm for a Drude aluminum nanoparticle. That analysis, however, evaluated the required surface integrals of the Maxwell stress tensor by adaptive numerical cubature over the polar and azimuthal angles at each frequency, a strategy whose cost grows quickly with multipole order, since the integrands involve increasingly oscillatory products of high-order associated Legendre functions. As a consequence, converged results at $a=50$~nm were reported only for a single Drude-like material at $\ell_{\max}=13$, gold and bismuth were examined only up to $a=5$~nm, and in every case the frequency-integrated angular momentum transfer $\Delta L$ was reported directly, without resolving or interpreting its spectral density. A complementary study within the dipolar approximation addressed the AMT to non-spherical NPs, specifically spheroids and Platonic solids, and showed that nanoparticle geometry strongly modulates the spectral weight of plasmonic resonances and thereby the magnitude of the transferred torque \cite{briseno2024angular}. The framework of that study relied on analytically or numerically known electric polarizabilities \cite{briseno2024angular} and remained restricted to the small-particle limit.

The present work closes this gap by eliminating the cubature bottleneck that restricted converged results to $\ell_{\max}=13$ at $a=50$~nm for aluminum alone, left gold and bismuth confined to $a=5$~nm, and yielded only the frequency-integrated $\Delta L$ rather than its spectral density. Building on the causal, multipole-convergent Maxwell stress tensor methodology developed for the linear-momentum problem in Ref.~\cite{castrejon2026electrodynamics}, and on the physical picture of angular momentum transfer established in Ref.~\cite{castellanos2023theory}, we show that the angular-momentum surface integrals admit the same analytical reduction: rather than being evaluated by numerical cubature, they collapse onto the same family of irreducible one-dimensional integrals over associated Legendre functions that underlies the linear-momentum calculation. This eliminates angular quadrature as the computational bottleneck and reduces the cost of the double multipolar sum from the naive $O(\ell_{\max}^4)$ to $O(\ell_{\max}^3)$ (Sec.~\ref{th C}). As a result, we push the multipole truncation to $\ell_{\max}=51$ for nanoparticles as large as $a=50$~nm, nearly four times the $\ell_{\max}=13$ reached in Ref.~\cite{castellanos2023theory} at the same particle size, and do so simultaneously for an optically simple (Drude \cite{Markovic}) and an optically complex, interband-dominated (Werner Au \cite{werner}) material, at a wall-clock computing time three to four orders of magnitude lower than reported for comparable multipole orders in Ref.~\cite{castellanos2021phdthesis} (Appendix~\ref{app: multipole convergence}).

Beyond this computational advance, resolving the full spectral density $\mathcal{L}(\omega)$, rather than only its frequency integral $\Delta L$, identifies which plasmonic and interband resonances actually carry the transferred torque. This lets us test, at nanoparticle sizes relevant to electron-tweezer experiments, whether the qualitative picture established in the small-particle regime, an electric-dominated angular momentum transfer driven by the interaction term, survives full retardation and high-order multipolar coupling for a material with a markedly more complex dielectric response than aluminum.

We apply the resulting method to aluminum and gold NPs with radii up to $a = 50$~nm, resolving the spectral density of angular momentum transfer across the full frequency domain and quantifying its dependence on nanoparticle size, electron speed, impact parameter, and material-specific plasmonic and interband resonances. Aluminum provides a canonical Drude-like reference, while gold, whose interband transitions peak near $4$~eV in optical measurements~\cite{johnsonchristy1972} and are superimposed on a plasmonic free-electron response, tests the framework under conditions of strong interband activity typical of noble metals. Throughout, causality of the dielectric input is enforced, and convergence with respect to $\ell_{\max}$ is explicitly verified. Our results establish a numerically controlled reference for angular momentum transfer in the large-particle regime and delineate the physical ingredients captured within an isolated, local, spherical electrodynamic description, providing a foundation for future extensions to non-spherical geometries, nanoparticle assemblies, and substrate-mediated interactions relevant to practical electron-tweezer implementations.

%%%%%%%%%%%%%%%%%%%%%%%%%%%%%%%%%%%%%%%%%%%%%%%%%%%%%%%%%%%%%%%%%%%%%%%%%%%%%%%%%
\section{Analytical expressions for the angular momentum transfer}
\label{th}
%%%%%%%%%%%%%%%%%%%%%%%%%%%%%%%%%%%%%%%%%%%%%%%%%%%%%%%%%%%%%%%%%%%%%%%%%%%%%%%%%

\begin{figure}[h!]
    \centering
    \includegraphics[width=0.8\linewidth]{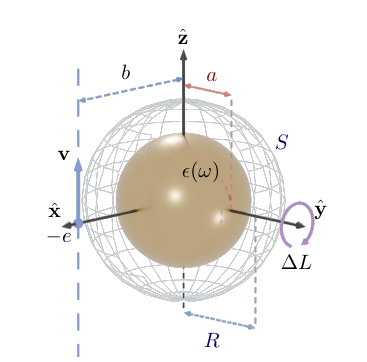}
    \caption{Schematics of the system under study. A spherical nanoparticle of radius $a$, characterized by a frequency-dependent dielectric function $\epsilon(\omega)$, is embedded in vacuum and centered at the origin. A swift electron (blue dot) travels with constant speed $\vb{v}=v\,\hb{z}$ at an impact parameter $b$ relative to the NP center. The angular-momentum transfer is obtained from a closed-surface integral of the Maxwell stress tensor over the integration surface $S$ (grey), a sphere of radius $R$ satisfying $a<R<b$, so that it fully encloses the nanoparticle without intersecting the electron trajectory; one wedge of $S$ facing the reader has been omitted purely for visual clarity.}
    \label{fig:system}
\end{figure}

We adopt the same local, fully causal electrodynamical framework used for the linear-momentum problem in Ref.~\cite{castrejon2026electrodynamics}: an uncharged, nonmagnetic spherical nanoparticle of radius $a$ and local dielectric function $\epsilon(\omega)$, embedded in vacuum, interacting with a swift electron of charge $-e$ traveling at constant speed $\vb{v}=v\,\hat{\vb{z}}$ at impact parameter $b$, as shown in Fig.~\ref{fig:system}. The total electromagnetic field is expressed as the superposition of the external field of the bare electron and the field scattered by the nanoparticle, each expanded in the multipole series of Ref.~\cite{castrejon2026electrodynamics}, Eqs.~(1)-(2), reproduced in Appendix~\ref{app: multipole expansion} for completeness. We summarize below only the elements specific to the angular-momentum; the reader is referred to Ref.~\cite{castrejon2026electrodynamics} for the shared field-expansion and stress-tensor formalism.

The total angular momentum transferred from the swift electron to the NP is obtained from the conservation of angular momentum in electrodynamics. In the frequency domain, the angular momentum transfer along an arbitrary direction $\hb{n}$ can be written as \cite{castellanos2023theory}
\begin{equation}
    \Delta{\mathbf L}\cdot\hb{n}=\Delta L_n = \int_0^{\infty} \mathcal{L}_n(\omega)\,d\omega,
\end{equation}
where $\mathcal{L}_n(\omega)$ is the spectral density of the angular momentum transfer.

By symmetry with respect to the $xz$ plane ($y=0$), the $y$ component of linear momentum vanishes. The same mirror symmetry independently constrains the angular momentum transfer: under $y\to-y$, $\vb{L}$ transforms as a pseudovector, so $L_x\to-L_x$ and $L_z\to-L_z$ while $L_y\to L_y$, leaving $\Delta L_y$ as the only nonvanishing component. Therefore, in the context of nanoparticle manipulation, $\Delta L_y$ is of primary interest, as it produces the rotational torque acting on the nanoparticle. We define $\Delta L\equiv\Delta L_y$ as the ordinary Cartesian $y$-component of $\vb{r}\times\vb{p}$ in the right-handed coordinate frame of Fig.~\ref{fig:system}; with this convention, $\Delta L>0$ corresponds to clockwise rotation and $\Delta L<0$ to counterclockwise rotation about the $y$ axis, as viewed in Fig.~\ref{fig:system}. The rotation arrow drawn in Fig.~\ref{fig:system} is oriented counterclockwise because this is the sign consistently obtained from our simulations across the parameter range explored, $\Delta L<0$ throughout, rather than an independent choice of convention.

%%%%%%%%%%%%%%%%%%%%%%%%%%%%%%%%%%%%%%%%%%%%%%%%%%%%%%%%%%%%%%%%%%%%%%%%%%%%%%%%%
\subsection{The spectral density of the angular momentum}\label{th A}
%%%%%%%%%%%%%%%%%%%%%%%%%%%%%%%%%%%%%%%%%%%%%%%%%%%%%%%%%%%%%%%%%%%%%%%%%%%%%%%%%
The spectral density of the angular momentum transfer is obtained from a closed-surface integral of the Maxwell stress tensor expressed in the frequency domain \cite{castrejon2021time}. We choose the integration surface $S$ as a sphere of radius $R$, centered at the origin, which fully encloses the nanoparticle and does not intersect the electron trajectory. The spectral density of the momentum transfer is then given by \cite{castellanos2023theory}
\begin{equation}
    \mathcal{L}_{i} (\omega) = \frac{1}{\pi} \oint_S \epsilon_{i}^{\,\,\,\, lj} r_l \mathscr{T}_{jk} (\vec{r},\omega) \, dS^{k},
    \label{eq: spectral density AMT}
\end{equation}
where Einstein summation convention has been used, and $\ten{\mathscr{T}}(\vb{r},\omega)$ is the Maxwell stress tensor in the frequency domain, defined as \cite{castrejon2021time}
\begin{align}
& \ten{\mathscr{T}} (\vb{r},\omega) = \, \, \!\Re \bigg\{\!
\epsilon_0 \Big[\vb{E}(\vb{r};\omega)\vb{E}^{*}(\vb{r};\omega) \!-\! \frac{\ten{I}}{2}\vb{E}(\vb{r};\omega) \! \cdot \!\vb{E}^{*}(\vb{r};\omega) \Big] \nonumber \\
& \,\, + \, \mu_0 \Big[\vb{H}(\vb{r};\omega)\vb{H}^{*}(\vb{r};\omega) \! - \! \frac{\ten{I}}{2}\vb{H}(\vb{r};\omega)\! \cdot \!\vb{H}^{*}(\vb{r};\omega) \Big] \! \bigg\}.
\label{eq: maxwell tensor freq}
\end{align}
Here, $\vb{E}$ and $\vb{H}$ denote the total electric and magnetic fields, respectively; $\Re\{\cdot\}$ indicates the real part; $(\cdot)^{*}$ indicates complex conjugation; and $\ten{I}$ is the unit dyadic. The stress tensor naturally separates into electric and magnetic contributions, each of which can be further decomposed into terms arising from the external field, the scattered field, and their interaction. Here, ``electric'' and ``magnetic'' contributions denote the $\epsilon_0[\cdots]$ and $\mu_0[\cdots]$ parts of the Maxwell stress tensor in Eq.~\eqref{eq: maxwell tensor freq}, evaluated from the total fields.

Accordingly, the electric contribution to the stress tensor can be written as
\begin{equation} \label{Tsum}
    \ten{\mathscr{T}}_{\rm E} = \ten{\mathscr{T}}^{\rm ee}_{\rm E}+\ten{\mathscr{T}}^{\rm ss}_{\rm E}+\ten{\mathscr{T}}^{\rm int}_{\rm E},
\end{equation}
with the interaction term defined as
\begin{equation}\label{Tint}
    \ten{\mathscr{T}}^{\rm int}_{\rm E} = \ten{\mathscr{T}}^{\rm es}_{\rm E}+\ten{\mathscr{T}}^{\rm se}_{\rm E}.
\end{equation}
As in Ref.~\cite{castrejon2026electrodynamics}, the superscripts $\text{e}$ and $\text{s}$ denote external and scattered fields, and the interaction term $\ten{\mathscr{T}}^{\rm int}_{\rm E}$ [Eq.~\eqref{Tint}] captures the cross contribution between the electron's field and the field scattered by the nanoparticle. Analogous terms follow for the magnetic contribution; for simplicity, we restrict the discussion below to the electric contribution. We show in Sec.~\ref{results} that this interaction term likewise dominates the angular momentum transfer, consistent with the near-field, interference-driven picture established for the linear-momentum case. Because the nanoparticle is electrically neutral before and after the interaction, and the external field alone carries no information about the nanoparticle, the external-external contribution $\ten{\mathscr{T}}^{\rm ee}_{\rm E}$ vanishes identically once integrated over the closed surface $S$, exactly as in the linear-momentum case \cite{castellanos2023theory,castrejon2026electrodynamics}. We retain it in Eq.~\eqref{Tsum} only to keep the notation parallel to Ref.~\cite{castrejon2026electrodynamics}.

Substituting the multipole expansions of Eq.~\eqref{eq:ScatterdEField} into Eq.~\eqref{Tsum}, the radial component of the electric contribution to the stress tensor can be written as
\begin{align}\label{Tradial}
    \ten{\mathscr{T}}^{(\text{e,s})(\text{e}^{\prime},\text{s}^{\prime})}_{\rm E} &\cdot \hb{r} = \epsilon_0 \sum_{\ell, m}\sum_{\ell^\prime,m^\prime} \Re \Bigg\{ \mathcal{E}_{\ell^\prime, m^\prime}^{(\text{e}^{\prime},\text{s}^{\prime})r*} \Big( \mathcal{E}_{\ell, m}^{(\text{e,s})r} \hb{r} \nonumber \\
    &+ \mathcal{E}_{\ell, m}^{(\text{e,s})\theta} \hat{\boldsymbol{\theta}} + \mathcal{E}_{\ell, m}^{(\text{e,s})\varphi} \hat{\boldsymbol{\varphi}}\Big)
    - \frac{1}{2}\Big( \mathcal{E}_{\ell, m}^{(\text{e,s})r} \mathcal{E}_{\ell^{\prime}, m^{\prime}}^{(\text{e}^{\prime},\text{s}^{\prime})r*} \nonumber \\
    &+ \mathcal{E}_{\ell, m}^{(\text{e,s})\theta} \mathcal{E}_{\ell^{\prime}, m^{\prime}}^{(\text{e}^{\prime},\text{s}^{\prime})\theta*} + \mathcal{E}_{\ell, m}^{(\text{e,s})\varphi} \mathcal{E}_{\ell^{\prime}, m^{\prime}}^{(\text{e}^{\prime},\text{s}^{\prime})\varphi*} \Big) \hb{r} \Bigg\},
\end{align}
where the superscripts $(\text{e,s})$ and $(\text{e}^{\prime},\text{s}^{\prime})$ again denote external or scattered field contributions.

The products of the electromagnetic field components appearing in Eq.~\eqref{Tradial} can be evaluated analytically, and the resulting angular surface integrals in Eq.~\eqref{eq: spectral density AMT} can likewise be carried out analytically for every term, reducing to a small set of irreducible integrals over associated Legendre functions that depend only on the integers $\ell,\ell^\prime,m,m^\prime$ and are independent of the material or trajectory parameters. We work through a representative term explicitly, and collect the complete set of irreducible integrals, in Appendix~\ref{app: momentum spectral density}\,\footnote{
The numerical implementation used in this work is openly available on GitHub~\cite{AMTRepo}.
Details on compilation, execution, and data organization are provided in the repository documentation and in Appendix \ref{app:numerical}.}.

%%%%%%%%%%%%%%%%%%%%%%%%%%%%%%%%%%%%%%%%%%%%%%%%%%%%%%%%%%%%%%%%%%%%%%%%%%%%%%%%%
\subsection{Computational scaling}\label{th C}
%%%%%%%%%%%%%%%%%%%%%%%%%%%%%%%%%%%%%%%%%%%%%%%%%%%%%%%%%%%%%%%%%%%%%%%%%%%%%%%%%

A direct evaluation of the double sum over $(\ell,m)$ and $(\ell^\prime,m^\prime)$ in Eq.~\eqref{Tradial}, truncated at $\ell_{\max}$, involves $O(\ell_{\max}^4)$ terms, since both $(\ell,m)$ and $(\ell^\prime,m^\prime)$ range independently over $O(\ell_{\max}^2)$ values. This scaling makes a brute-force evaluation of the quadratic field products entering the stress tensor prohibitively expensive for the large nanoparticles relevant to electron-tweezer applications, and the cost is compounded further if the angular integral over the closed surface $S$ is itself carried out numerically: because the products $P_\ell^m P_{\ell^\prime}^{m^\prime}$ oscillate increasingly rapidly as $\ell,\ell^\prime$ grow, adaptive angular cubature requires an increasing number of sampling points to maintain a fixed target accuracy, further raising the effective cost per multipole pair.

Following the same strategy introduced for the linear-momentum stress tensor in Ref.~\cite{castrejon2026electrodynamics}, we avoid numerical angular quadrature altogether. As illustrated by the worked example in Appendix~\ref{app: momentum spectral density}, every surface integral appearing in Eq.~\eqref{eq: spectral density AMT} factorizes into an azimuthal integral $\int_0^{2\pi} \rme^{i(m-m^\prime)\varphi}\{\cos\varphi,\sin\varphi,1\}\,d\varphi$, which vanishes unless $m^\prime=m$ or $m^\prime=m\pm1$, and a polar integral over a product of associated Legendre functions and elementary trigonometric weights, reducible to the irreducible integrals $IM,IU,IV,IW$, and $\Delta_{\ell,\ell^\prime}^m$ defined in Appendix~\ref{app: momentum spectral density}. Because these integrals depend only on the integers $\ell,\ell^\prime,m,m^\prime$, they can be evaluated once, to machine precision, using Gauss-Legendre and Gauss-Chebyshev quadratures~\cite{kahaner1989numerical,press2007numerical}, independently of the material or trajectory parameters, and reused for every frequency, speed, and impact parameter considered.

The azimuthal selection rule $m^\prime\in\{m,m\pm1\}$ reduces the number of surviving $(\ell,m,\ell^\prime,m^\prime)$ quadruples from $O(\ell_{\max}^4)$ to $O(\ell_{\max}^3)$: for each of the $O(\ell_{\max}^2)$ pairs $(\ell,m)$, only $O(1)$ values of $m^\prime$ contribute, each admitting $O(\ell_{\max})$ allowed values of $\ell^\prime$. Together with standard recurrence relations for the spherical Bessel and Hankel functions $Z_\ell^{\rm (e,s)}$ and for the associated Legendre functions $P_\ell^m$~\cite{kahaner1989numerical,press2007numerical}, this reduction is what makes it possible to converge the multipole truncation to $\ell_{\max}=51$ for nanoparticles as large as $a=50$~nm, for both optically simple and optically complex materials, at a cost that scales more favorably than a direct numerical evaluation of the same surface integrals (see Appendix~\ref{app: multipole convergence} for the explicit convergence tests underlying this claim, and Fig.~\ref{fig:cost_scaling} there for a direct empirical test of the $O(\ell_{\max}^3)$ scaling itself).

The efficiency gain established above is specific to the spherical geometry considered here, and it is worth situating it explicitly against the general-purpose computational electrodynamics methods most commonly used to model nanoparticle optical and electron-beam response: the boundary-element method (BEM), widely implemented for nanophotonics in the MNPBEM toolbox~\cite{hohenester2012mnpbem} and applied to electron-beam spectroscopies since Ref.~\cite{garciadeabajo2002retarded}; the finite-difference time-domain method (FDTD)~\cite{taflove2005computational}; and the discrete-dipole approximation (DDA)~\cite{draine1988}, alongside the closely related T-matrix method, all reviewed in the specific context of metal nanoparticles in Ref.~\cite{myroshnychenko2008modelling}. These methods discretize either the nanoparticle surface (BEM), the surrounding volume (FDTD), or the nanoparticle itself (DDA) into a mesh or lattice whose element size controls the numerical error; convergence therefore requires refining this discretization for every nanoparticle size and, for a moving-charge source, resolving both the fine spatial structure of the swift electron's near field and the plasmonic response of the nanoparticle on the same grid, a requirement that becomes increasingly demanding as $a$ grows toward the electron-tweezer-relevant sizes considered here. For a sphere, by contrast, the multipole expansion is not an approximation but an exact, complete basis for the geometry, so the only convergence parameter is the scalar truncation order $\ell_{\max}$, systematically verified in Appendix~\ref{app: multipole convergence}, rather than a mesh or grid that must be re-validated for each new size, material, or trajectory. Moreover, because the irreducible integrals $IM,IU,IV,IW,\Delta_{\ell,\ell^\prime}^m$ depend only on $(\ell,\ell^\prime,m,m^\prime)$, they are computed once and reused across the entire speed, impact-parameter, and material sweep reported in Sec.~\ref{results}, whereas mesh-based solvers generally require a new simulation for each trajectory. The cost of this efficiency is specialization: the present approach applies to isolated spherical nanoparticles, and general-purpose methods such as BEM, FDTD, or DDA remain the appropriate, and in fact necessary, tools for the non-spherical, inhomogeneous, or multi-particle geometries beyond the scope of this work.

%%%%%%%%%%%%%%%%%%%%%%%%%%%%%%%%%%%%%%%%%%%%%%%%%%%%%%%%%%%%%%%%%%%%%%%%%%%%%%%%%
\section{Angular Momentum Transferred to Aluminum and Gold Nanoparticles}
\label{results}
%%%%%%%%%%%%%%%%%%%%%%%%%%%%%%%%%%%%%%%%%%%%%%%%%%%%%%%%%%%%%%%%%%%%%%%%%%%%%%%%%

\begin{figure*}
\centering
\includegraphics[width=.99\linewidth]{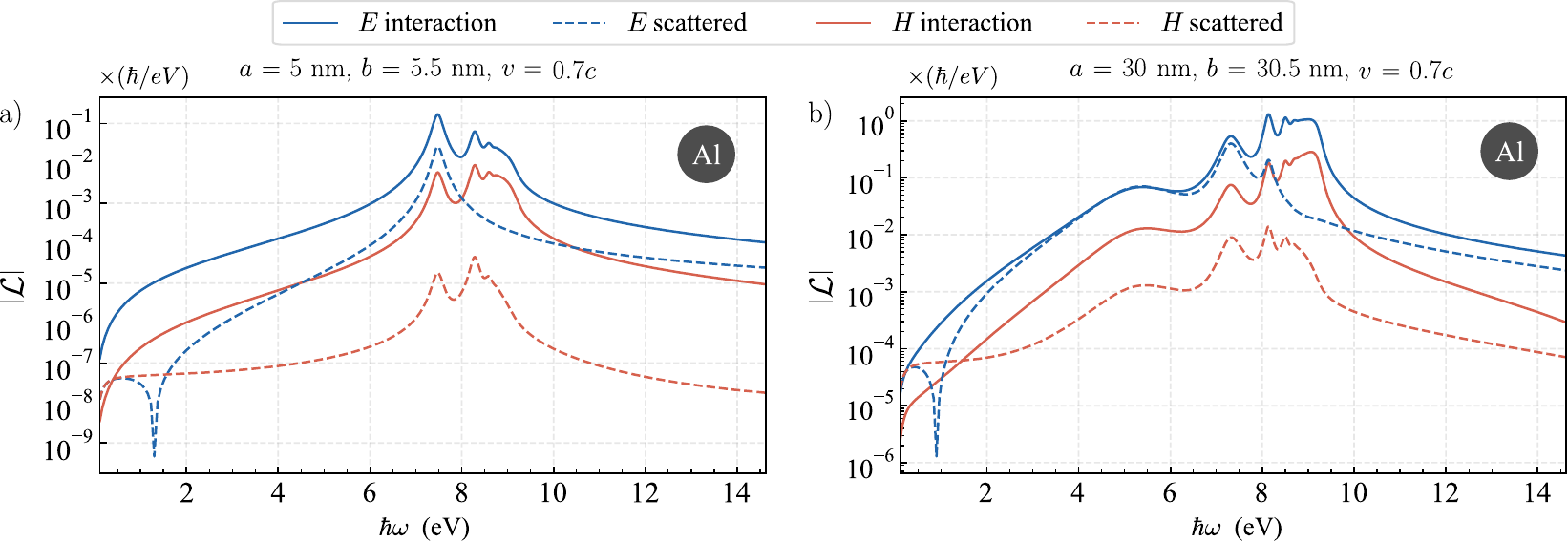}
    \caption{Frequency-resolved spectral density $|\mathcal{L}|$ in Log scale, decomposed into its electric (blue) and magnetic (red) contributions, each split into the interaction (solid) and scattered-scattered (dashed) terms of the Maxwell stress tensor [Eq.~\eqref{eq: spectral density AMT}]; the external-external term vanishes identically. a) Aluminum, $a=5$~nm, $b=5.50$~nm, $v=0.7c$. b) Aluminum, $a=30$~nm, $b=30.50$~nm, $v=0.7c$.}
\label{fig:spectral_density}
\end{figure*}

In this Section, we analyze the scalar angular momentum transfer from a swift electron to a spherical nanoparticle, $\Delta L \equiv \Delta L_y$, together with its spectral density $\mathcal{L}(\omega) \equiv \mathcal{L}_y(\omega)$, defined with respect to the geometry of Fig.~\ref{fig:system}. To extract the roles of retardation, multipolar order, and material response, we report results at several nanoparticle radii: the frequency-resolved spectral density is examined at $a=5$ and $30$~nm, where dense sampling of $\mathcal{L}(\omega)$ over the full frequency range remains computationally tractable, while the frequency-integrated $\Delta L$ is followed as a function of electron speed and impact parameter up to $a=50$~nm, the size regime most directly relevant to electron-tweezer manipulation, for both materials studied.

Accurate evaluation of the angular momentum transfer requires explicit convergence of the multipole expansions in Eqs.~\eqref{eq:ScatterdEField} and~\eqref{eq:ScatterdMField}; the minimum truncation order needed for a target accuracy grows with nanoparticle radius, as verified explicitly across the full size range studied in Appendix~\ref{app: multipole convergence}. For the production runs reported below we adopt, rather than the bare per-size minimum, a uniformly generous truncation, well beyond what a $1\%$ tolerance between successive orders requires: $\ell_{\max}=44$ for aluminum and $\ell_{\max}=51$ for gold at $a=30$~nm, and $\ell_{\max}=51$ for both materials at $a=50$~nm, the exact particle size and Drude parametrization for which Ref.~\cite{castellanos2023theory} reported $\ell_{\max}=13$, a comparison developed quantitatively in Sec.~\ref{sec:comparison}. For gold specifically, our results are converged at a particle size ten times larger than the $a=5$~nm limit to which Ref.~\cite{castellanos2023theory} treatment of optically complex materials was restricted. The analytical expressions derived in Sec.~\ref{th} allow $\mathcal{L}(\omega)$ to be evaluated to machine precision, and the subsequent frequency integration yields $\Delta L$. The frequency cutoff is chosen so that the neglected tail changes $\Delta L$ by less than $10^{-4}$ in relative terms. The multipole order is increased until the relative change of $\Delta L$ between successive orders falls below $10^{-4}$, or up to $\ell_{\max}=51$. The latter is reached only for electron trajectories passing within a few nanometers of the nanoparticle surface at $a\geq30$~nm. There, the final change is at most $8\times10^{-4}$ for aluminum and a few times $10^{-3}$ for gold. Since $\Delta L$ increases monotonically with $\ell_{\max}$, the truncated values underestimate $|\Delta L|$. This underestimate is largest for trajectories passing closest to the surface, but it remains below the resolution of the plots and does not affect any of the trends discussed below. The convergence behavior is illustrated in Appendix~\ref{app: multipole convergence}.

We consider two materials with markedly different dielectric responses. Aluminum is modeled using a Drude dielectric function \cite{Markovic} and serves as a canonical plasmonic metal. Gold is described by the causal dielectric function of Werner \emph{et al.}~\cite{werner}, fitted to a Drude term plus eight Lorentz oscillators, capturing both its free-electron plasmonic response and its well-known interband transitions; the parameters of this fit are reported in Appendix~\ref{app: dielectric Au}. Gold, rather than the bismuth used as the optically complex material in the companion linear-momentum-transfer study~\cite{castrejon2026electrodynamics}, is chosen here because it is both the more widely used plasmonic material in electron-microscopy and nanophotonics applications, and it is the material for which electron-beam-induced nanoparticle motion, including translational displacement and reorientation, has been directly documented experimentally~\cite{Batson,Batson2,Batson3}, making it the natural optically complex material against which to benchmark the angular momentum transfer computed here. This choice allows us to contrast angular momentum transfer in an optically simple and an optically complex material while maintaining a fully causal description throughout. Size corrections to the dielectric function are neglected, following Ref.~\cite{castellanos2023theory}, which showed such corrections to be negligible for the momentum-transfer problem at the nanoparticle radii considered here.

Details of the numerical implementation, as well as additional dielectric functions already available in the code, are documented in the GitHub repository~\cite{AMTRepo}, with a brief summary provided in Appendix~\ref{app:numerical}.

%%%%%%%%%%%%%%%%%%%%%%%%%%%%%%%%%%%%%%%%%%%%%%%%%%%%%%%%
\subsection{Angular momentum transferred to a Drude-like aluminum nanoparticle}
%%%%%%%%%%%%%%%%%%%%%%%%%%%%%%%%%%%%%%%%%%%%%%%%%%%%%%%%

We first analyze the angular momentum transferred to an aluminum nanoparticle modeled by a Drude dielectric function with parameters $\hbar\omega_p = 13.14$~eV and $\hbar\Gamma = 0.197$~eV~\cite{Markovic}. Because of its simple free-electron response, aluminum provides a convenient reference system for studying the roles of multipolar resonances, electron speed, and impact parameter in the momentum-transfer process.

Figures~\ref{fig:spectral_density}a) and~\ref{fig:spectral_density}b) show the spectral density of the angular momentum transfer, $\mathcal{L}(\omega)$, for a representative swift electron incident on an aluminum nanoparticle of radius $a=5$ and $30$~nm, respectively, at fixed speed $v=0.7c$ and impact parameter $b=a+0.5$~nm. Both spectra are dominated by a cluster of resonances between $5$ and $9$~eV, below the asymptotic dipolar surface-plasmon frequency $\hbar\omega_s=\hbar\omega_p/\sqrt{2}\approx9.3$~eV~\cite{Bohren}. Retardation redshifts the plasmonic response and, for the larger nanoparticle, splits it into several closely spaced multipolar resonances: these are barely resolved at $a=5$~nm but clearly separated into three sub-peaks spanning $8$-$9$~eV at $a=30$~nm. The electric-interaction term exceeds the magnetic-interaction term at every frequency shown, confirming, now at the level of the spectral density rather than only its frequency integral, the electric-dominated picture established in Ref.~\cite{castellanos2023theory}; the margin, however, is only a factor of $\approx20$ at $a=5$~nm and $\approx5$-$7$ at $a=30$~nm over most of the spectrum, and narrows to $\approx4$ near the upper edge of the cluster of resonances ($\hbar\omega\approx9$~eV), so that the magnetic channel is subdominant but not negligible.

The interaction (solid) and scattered-scattered (dashed) contributions in Figs.~\ref{fig:spectral_density}a) and \ref{fig:spectral_density}b) reveal a more subtle picture than a simple hierarchy of magnitudes. Because $|\mathcal{L}|$ is plotted on a logarithmic scale, a sign change appears as a sharp downward cusp. The electric-interaction term of $\mathcal{L}$ (solid blue line in Fig.~\ref{fig:spectral_density}) retains a single sign across the entire spectrum shown in both panels of Fig.~\ref{fig:spectral_density}, but the electric-scattered term of $\mathcal{L}$ (dashed blue line in Fig.~\ref{fig:spectral_density}) shows one isolated sign change at low frequency, well below the main cluster of resonances ($\sim\!1$~eV at $a=5$~nm, shifting slightly with size), where it is several orders of magnitude smaller than the interaction term and therefore does not influence the sign of the total. Comparison with the frequency-integrated electric components in Fig.~\ref{fig:DL_v_Al} below shows that, across the cluster of resonances that dominates the frequency integral, the interaction and scattered-scattered terms carry opposite signs, with the larger magnitude of the interaction term (by roughly one order of magnitude at $a=5$~nm, narrowing to a factor of a few at $a=30$~nm) setting the sign of the total. The magnetic channel shows no sign change within the spectral range shown: both the magnetic-interaction $\mathcal{L}$ (solid red line in Fig.~\ref{fig:spectral_density}) and magnetic-scattered $\mathcal{L}$ (dashed red line in Fig.~\ref{fig:spectral_density}) retain a single sign across the entire spectrum shown in both panels (the magnetic-interaction term changes sign only above it, near $23$~eV at $a=5$~nm and $17$~eV at $a=30$~nm, where $|\mathcal{L}|$ is three to four orders of magnitude below its peak), the interaction term exceeding the scattered-scattered one by two to three orders of magnitude at $a=5$~nm and one to two at $a=30$~nm across the cluster of resonances and closely tracking the resonance structure of the electric-interaction term. The isolated low-frequency sign change of the electric-scattered term is thus the only sign change of an individual term within the spectral range shown; it is invisible to a frequency-integrated calculation but, carrying negligible spectral weight, does not affect the sign of any of the integrated contributions discussed below.

\begin{figure}
\centering
\includegraphics[width=.85\linewidth]{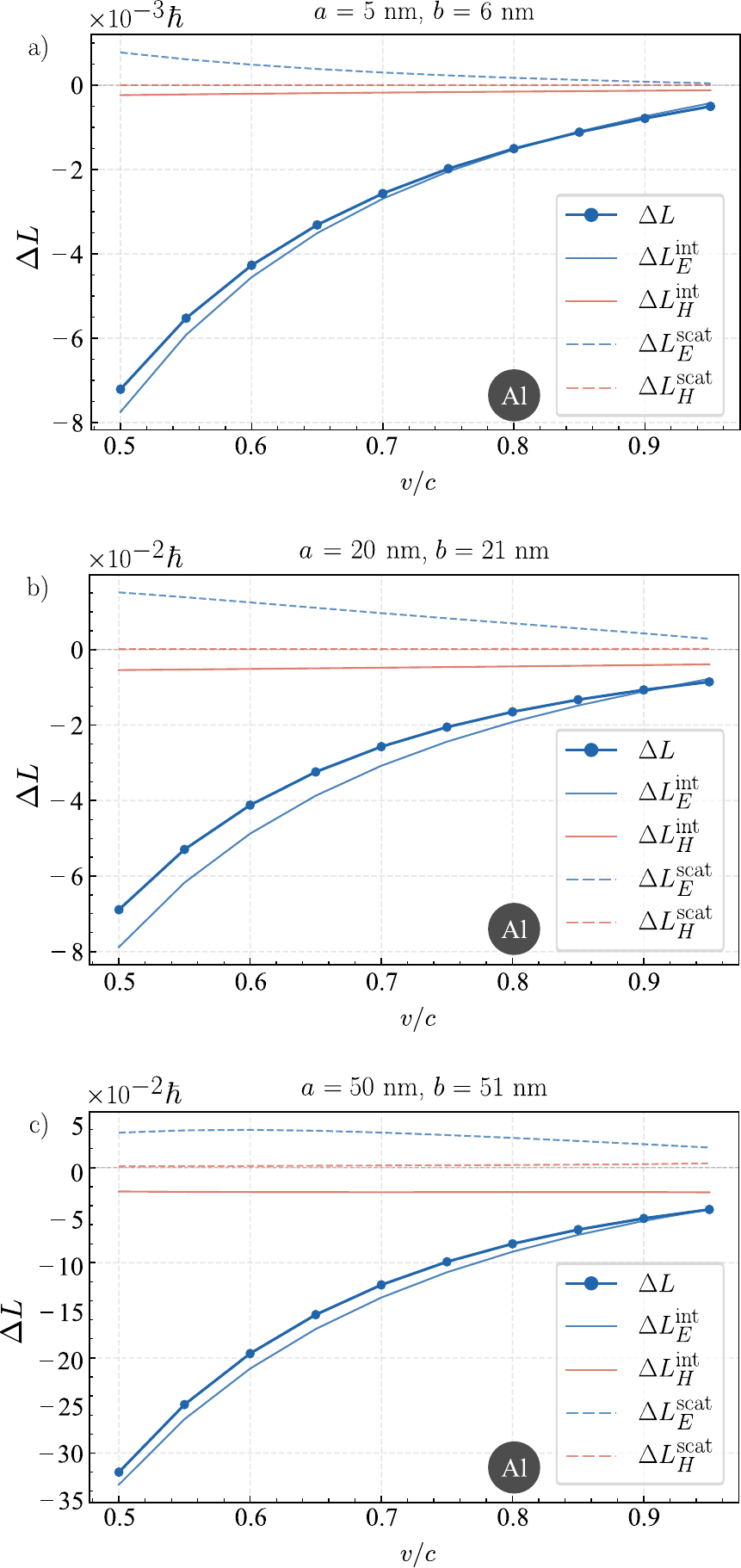}
    \caption{Total angular momentum transfer $\Delta L$ (thick line with markers), and its electric interaction ($\Delta L_E^{\rm int}$, solid blue line), electric scattered-scattered ($\Delta L_E^{\rm scat}$, dashed blue line), magnetic interaction ($\Delta L_H^{\rm int}$, solid red line), and magnetic scattered-scattered ($\Delta L_H^{\rm scat}$, dashed red line) contributions, transferred by a swift electron to an aluminum NP as a function of $v$, at fixed $b=a+1$~nm: a) $a=5$~nm, $b=6$~nm; b) $a=20$~nm, $b=21$~nm; c) $a=50$~nm, $b=51$~nm.}
    \label{fig:DL_v_Al}
\end{figure}

\begin{figure*}
\centering
\includegraphics[width=.99\linewidth]{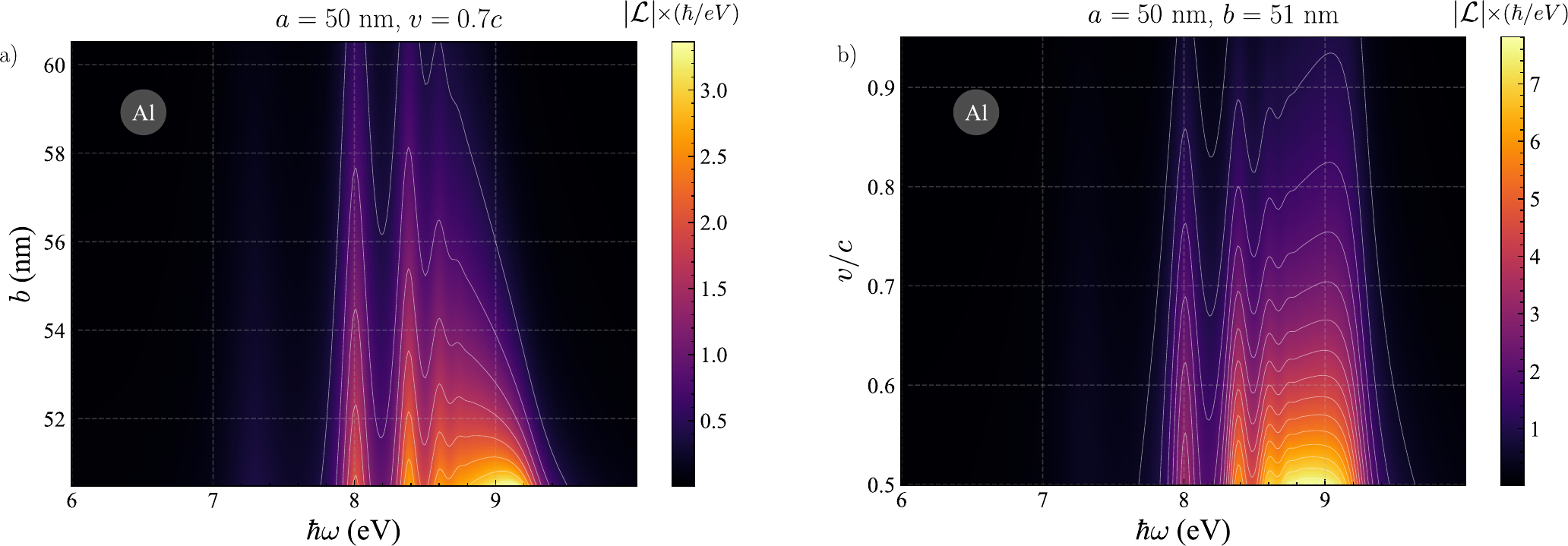}
    \caption{Frequency- and parameter-resolved maps of $|\mathcal{L}|$ at $a=50$~nm. a) Aluminum, as a function of $\hbar\omega$ and impact parameter $b$ at fixed $v=0.7c$. b) Aluminum, as a function of $\hbar\omega$ and electron speed $v$ at fixed $b=51$~nm.}
\label{fig:maps}
\end{figure*}

Figures~\ref{fig:maps}a) and~\ref{fig:maps}b) map the $7.5$-$9.5$~eV cluster of multipolar resonances at $a=50$~nm, as a function of impact parameter at fixed speed and of speed at fixed impact parameter, respectively. The spectral density decays rapidly with increasing $b$ at fixed $v$ [panel a)], consistent with the near-field origin of the interaction and with the analogous decay reported for the linear-momentum spectral density in Ref.~\cite{castrejon2026electrodynamics}. At fixed $b$, the peak spectral density instead decreases monotonically with increasing $v$ [panel b)], falling by a factor of $8.2$ between $v=0.5c$ and $v=0.95c$, comparable to the $7.3$-fold decrease of the frequency-integrated $|\Delta L|$ shown in Fig.~\ref{fig:DL_v_Al}c) over the same range. The resonant response and its frequency integral thus weaken together, at comparable rates, rather than standing in tension with one another. The dominant sub-resonance is the highest-lying one at every speed, drifting from $8.90$~eV at $v=0.5c$ to $9.04$~eV at $v=0.95c$, while the sub-resonance near $8.38$~eV comes within $4\%$ of it at $v\approx0.67c$ without overtaking it. The electric contribution alone would be dominated by that lower sub-resonance for $0.60c\lesssim v\lesssim0.82c$; it is the magnetic contribution, largest relative to the electric one near the upper edge of the cluster [Fig.~\ref{fig:spectral_density}], that keeps the uppermost sub-resonance dominant throughout, a direct spectral fingerprint of the closely spaced multipolar sub-clusters noted above.

We compare the total transverse angular momentum transfer $\Delta L$ as a function of $v$ at three nanoparticle radii: $a=5$, $20$, and $50$~nm, together with its electric and magnetic, interaction and scattered-scattered contributions, in Fig.~\ref{fig:DL_v_Al}. At every size, $\Delta L$ is negative (counterclockwise in Fig.~\ref{fig:system}) and decreases monotonically in magnitude with increasing $v$; its magnitude also grows monotonically with nanoparticle radius, from $\Delta L \approx -0.0072 \hbar$ at $a=5$~nm to $\Delta L \approx -0.320 \hbar$ at $a=50$~nm, at $v=0.5c$. The electric interaction term is the largest individual contribution at every size and speed, while the electric scattered-scattered term is smaller in magnitude and positive, partially canceling it. The magnetic contributions follow the same pattern, with a negative interaction term and a much smaller, positive scattered-scattered term, and, although smaller than the electric ones, are far from negligible: $|\Delta L_H^{\rm int}|$ varies far less with $v$ than the electric terms (it decreases by a factor of $2.0$ at $a=5$~nm and $1.4$ at $a=20$~nm and is nearly constant at $a=50$~nm, whereas $|\Delta L_E^{\rm int}|$ falls by factors of $18$, $10$, and $7.6$), so that its share of the total grows steadily as the electric terms decay with increasing $v$. The magnetic contribution $\Delta L_H=\Delta L_H^{\rm int}+\Delta L_H^{\rm scat}$ amounts to only $3$-$8\%$ of $\Delta L$ at $v=0.5c$ across the three sizes, but at $v=0.95c$ it reaches $24\%$ of $\Delta L$ at $a=5$~nm and $43\%$ and $49\%$ of it at $a=20$ and $50$~nm, respectively, where the electric and magnetic contributions become comparable [Fig.~\ref{fig:DL_v_Al}(c)]. Ref.~\cite{castellanos2023theory} reports that the magnetic contribution never changes sign in $\Delta L$ for aluminum up to $a=50$~nm, or for gold and bismuth up to the $a=5$~nm limit reached in that work. The present, more fully converged and spectrally resolved calculation confirms this at every size studied: neither $\Delta L_H^{\rm int}$ nor the total magnetic contribution $\Delta L_H$ changes sign anywhere in the range $0.5c\leq v\leq0.95c$ explored, so that the magnetic contribution reinforces the electric one throughout, consistent with the single-signed magnetic-interaction spectral density of Fig.~\ref{fig:spectral_density}.

%%%%%%%%%%%%%%%%%%%%%%%%%%%%%%%%%%%%%%%%%%%%%%%%%%%%%%%%%%%%%%%%%
\subsection{Angular momentum transferred to a gold nanoparticle}
%%%%%%%%%%%%%%%%%%%%%%%%%%%%%%%%%%%%%%%%%%%%%%%%%%%%%%%%%%%%%%%%%
\begin{figure}
\centering
\includegraphics[width=0.9\linewidth]{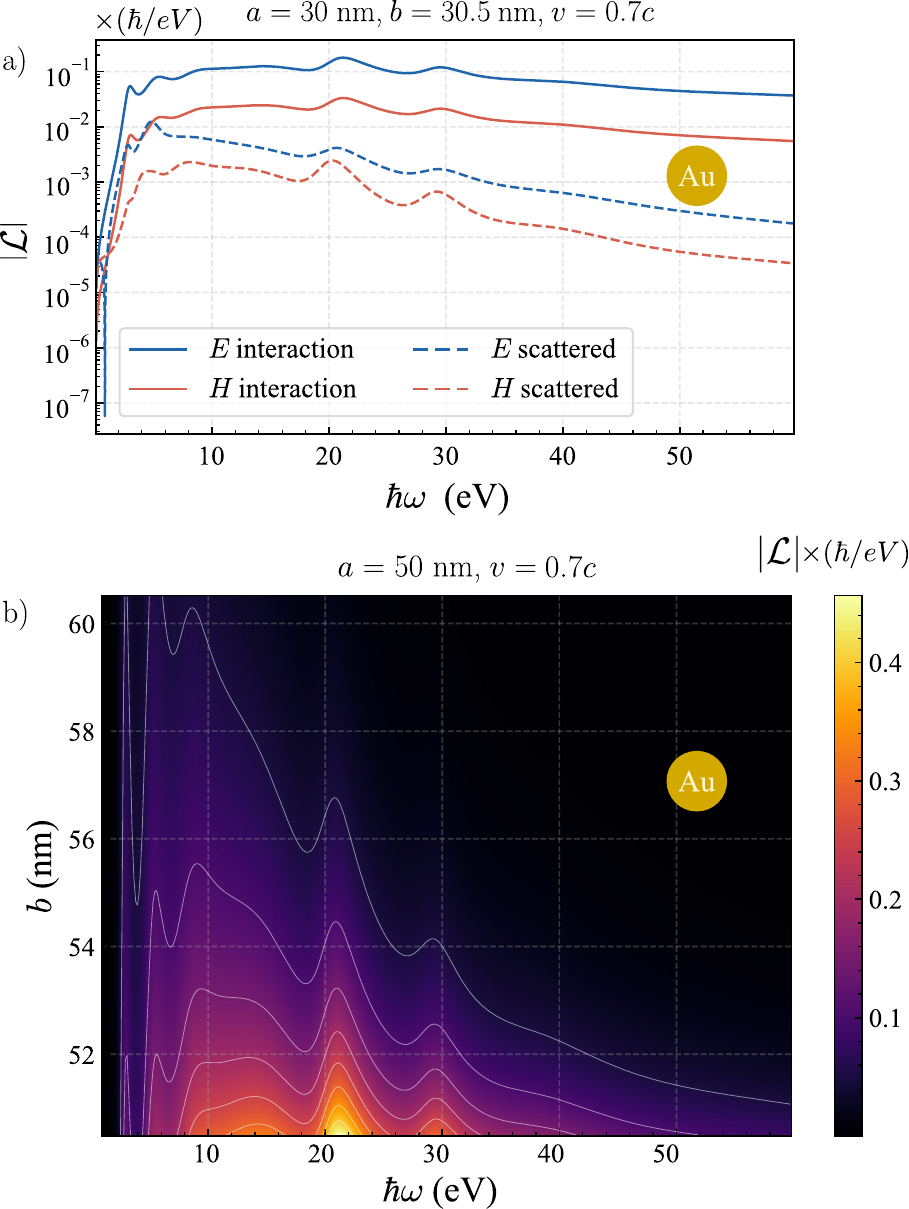}
    \caption{a) Frequency-resolved angular momentum transfer to a gold nanoparticle of radius $a=30$~nm, at fixed $v=0.7c$.  Spectral density $|\mathcal{L}(\omega)|$ in Log scale at $b=30.5$~nm, decomposed into its electric interaction ($E$ interaction, solid blue line), electric scattered-scattered ($E$ scattered, dashed blue line), magnetic interaction ($H$ interaction, solid red line), and magnetic scattered-scattered ($H$ scattered, dashed red line) contributions. b) Frequency- and impact-parameter-resolved map of $|\mathcal{L}|$ for $a=50$~nm, at fixed $v=0.7c$, over $50.5~\mathrm{nm}\leq b\leq60.5$~nm.}
    \label{fig:maps_Au30}
\end{figure}

Gold is modeled using the bulk dielectric function reported by Werner \emph{et al.}~\cite{werner}, expressed as a superposition of one Drude term and eight Lorentz oscillators accounting for its interband transitions. The corresponding parameters are provided in Appendix~\ref{app: dielectric Au}.

Figure~\ref{fig:maps_Au30}a) shows the spectral density of the angular momentum transfer for a gold nanoparticle of radius $a=30$~nm, at the same trajectory as in Fig.~\ref{fig:spectral_density}b). In marked contrast to the single cluster of resonances of aluminum, the electric-interaction spectral density for gold forms a broad, structured plateau extending from approximately $5$ to $40$~eV, with modest local maxima near $5$-$6$, $20$-$22$, and $30$~eV, before decaying above $\sim\!40$~eV. This reflects the superposition of gold's plasmonic free-electron response with its interband transitions, which, unlike aluminum's single Drude pole, distribute oscillator strength over a much wider energy range~\cite{werner} (see Appendix~\ref{app: dielectric Au}). The electric contribution again dominates the magnetic one throughout, and the interaction term dominates the scattered-scattered term in both channels, as for aluminum; for gold, however, the magnetic-interaction term (solid red line in Fig.~\ref{fig:maps_Au30}a)) is the second-largest contribution, closely following the structure of the electric-interaction plateau at a nearly constant factor of $\approx5$-$7$ below it above $\approx3$~eV, and exceeding the electric-scattered term above $\approx5$~eV. The electric-scattered term shows its own isolated low-frequency sign change, near $0.8$~eV, well below the plateau and orders of magnitude smaller than the electric-interaction term, whereas neither magnetic term changes sign anywhere in the spectrum shown. %fig:maps_Au30

Figure~\ref{fig:maps_Au30}b) maps the spectral density for a gold NP of radius $a=50$~nm as a function of $\hbar\omega$ and impact parameter at fixed $v=0.7c$, showing that the multi-peak structure persists and decays with $b$ over the full $0$-$60$~eV range, with the strongest weight concentrated below $\sim\!30$~eV.
\begin{figure}
    \centering
\includegraphics[width=.85\linewidth]{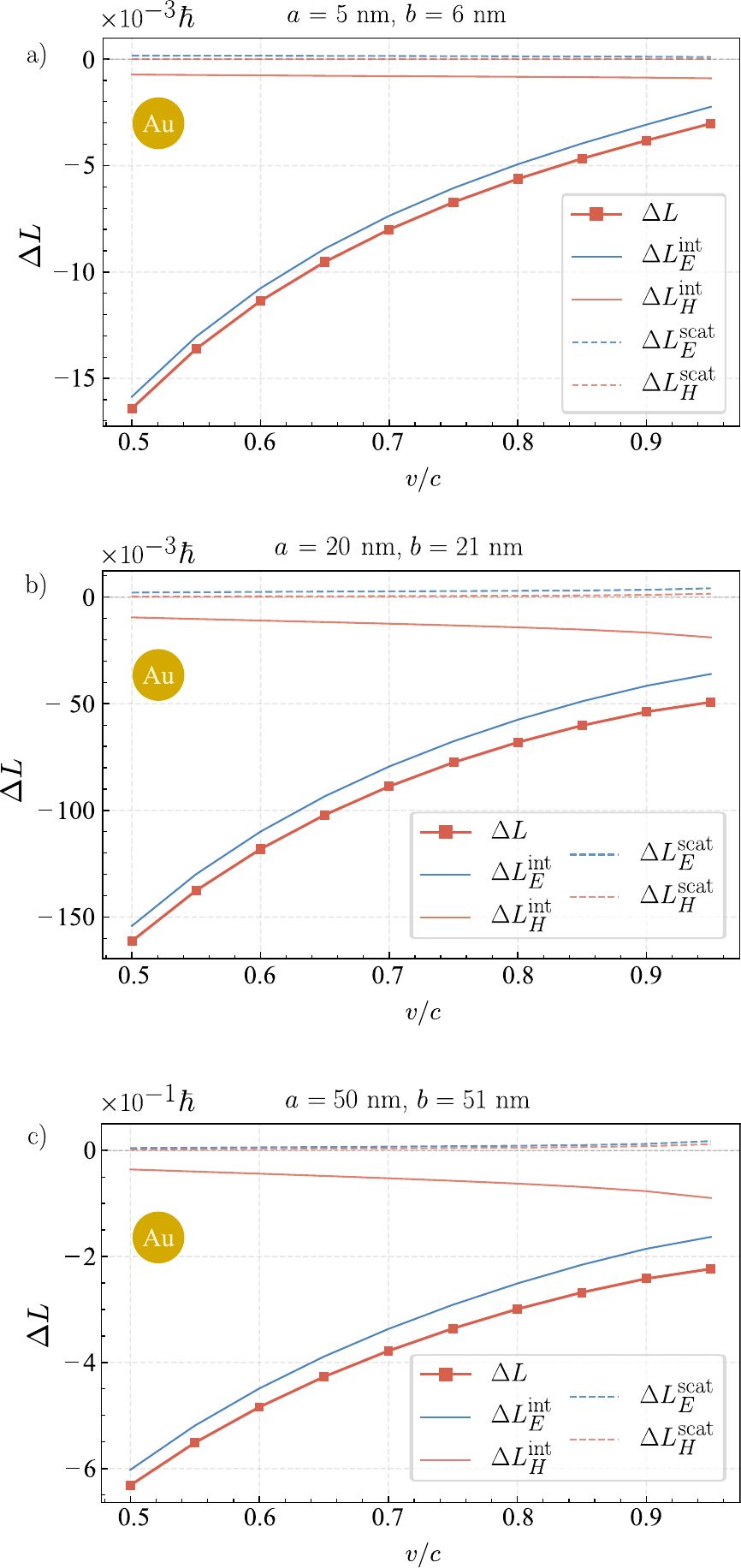}
    \caption{Total angular momentum transfer $\Delta L$ (solid red line with markers), and its electric interaction ($\Delta L_E^{\rm int}$, solid blue line), electric scattered-scattered ($\Delta L_E^{\rm scat}$, dashed blue line), magnetic interaction ($\Delta L_H^{\rm int}$, solid red line), and magnetic scattered-scattered ($\Delta L_H^{\rm scat}$, dashed red line) contributions, transferred by a swift electron to a gold NP as a function of $v$, at fixed $b=a+1 $~nm, following the same panel layout and convention as Fig.~\ref{fig:DL_v_Al}: a) $a=5$~nm, $b=6$~nm; b) $a=20$~nm, $b=21$~nm; c) $a=50$~nm, $b=51$~nm.}
    \label{fig:DL_v_Au}
\end{figure}

\begin{figure*}
    \centering
    \includegraphics[width=0.9\linewidth]{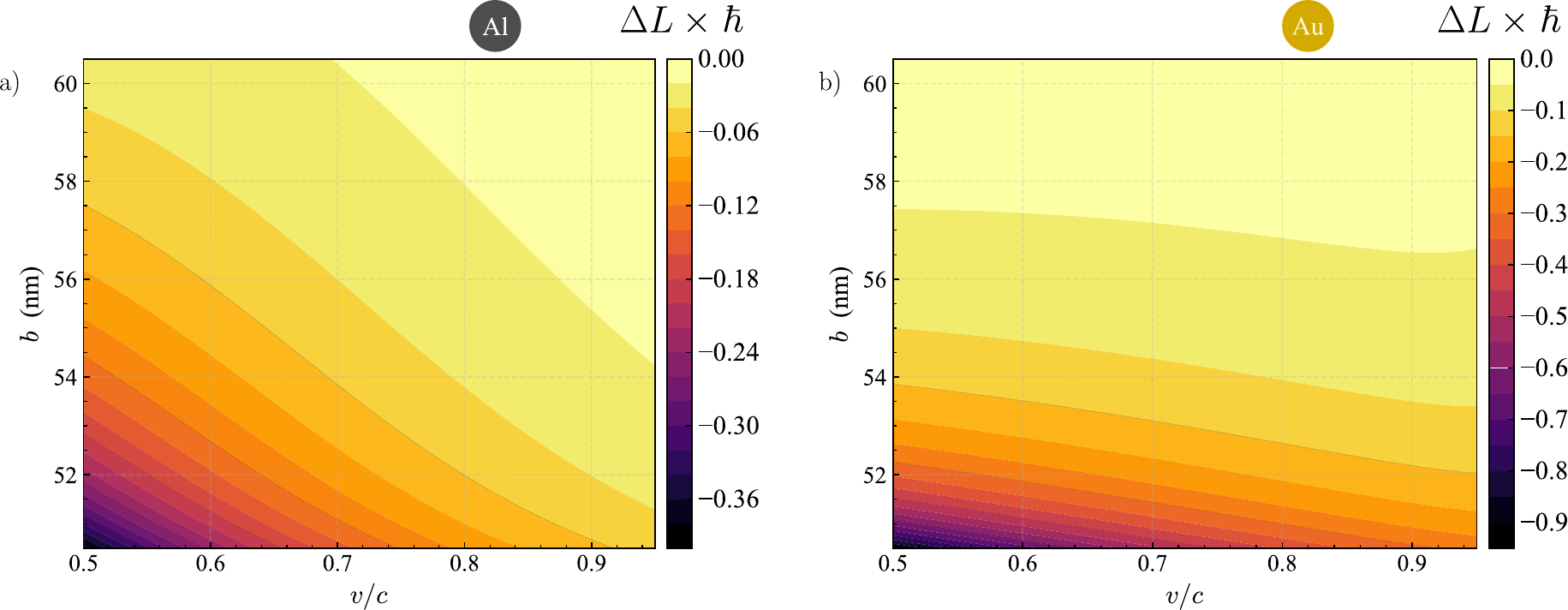}
    \caption{Frequency-integrated angular momentum transfer $\Delta L$ as a function of electron speed $v$ and impact parameter $b$, for a nanoparticle of radius $a=50$~nm: a) aluminum and b) gold. Color encodes $\Delta L$, negative (counterclockwise in Fig.~\ref{fig:system}) throughout both panels; note the different color scales.}
    \label{fig:Al_vs_Au}
\end{figure*}

Figure~\ref{fig:DL_v_Au} shows the gold analog of Fig.~\ref{fig:DL_v_Al}. The overall behavior parallels aluminum in several respects: $\Delta L$ is negative and decreases monotonically in magnitude with increasing $v$ at every size, and its magnitude also grows monotonically with nanoparticle radius, from $\Delta L \approx -0.0164 \hbar$ at $a=5$~nm to $\Delta L \approx -0.632 \hbar$ at $a=50$~nm, at $v=0.5c$, by a factor of $9.8$ from $a=5$ to $20$~nm and a further factor of $3.9$ from $a=20$ to $50$~nm. This closely tracks the aluminum size-scaling reported above ($9.6\times$ and $4.6\times$ over the same two steps), despite the markedly different dielectric responses of the two materials. The electric contribution again exceeds the magnetic one at every size and speed, consistent with the finding of Ref.~\cite{castellanos2023theory} for gold at $a=5$~nm and extending it to $a=50$~nm.

The electric channel of gold is quantitatively distinct from that of aluminum: the scattered-scattered term $\Delta L_E^{\rm scat}$ is smaller relative to $\Delta L_E^{\rm int}$ than for aluminum at every size, and by $a=50$~nm [panel c)] it amounts to only $1$-$11\%$ of $\Delta L_E^{\rm int}$, so that the electric contribution is dominated by its interaction term, $\Delta L_E\simeq\Delta L_E^{\rm int}$, in contrast to aluminum at the same size, where $\Delta L_E^{\rm scat}$ partially cancels $\Delta L_E^{\rm int}$ by a visibly larger margin [Fig.~\ref{fig:DL_v_Al}c)]. We attribute this weaker cancellation to the broader, more structured spectral density of gold [Fig.~\ref{fig:maps_Au30}a)], which distributes the scattered-scattered contribution over a much wider frequency range rather than concentrating it, in phase opposition to the interaction term, within the single narrow resonance that dominates aluminum's response. The magnetic terms remain a small fraction of the total at low speed for both materials, but as $v\to c$ and the electric terms decay, the magnetic contribution, dominated by an interaction term that varies far less with $v$ than the electric terms for aluminum and grows in magnitude with $v$ for gold, makes up a rapidly increasing share of $\Delta L$, reaching $35\%$ of the total for gold and $43$-$49\%$ of it for aluminum at $v=0.95c$ and $a=20$-$50$~nm.

The sign of the magnetic contribution is the same for gold as for aluminum. The scattered-scattered term $\Delta L_H^{\rm scat}$ remains positive throughout the range explored at all three sizes, while the interaction term $\Delta L_H^{\rm int}$ remains negative, and larger in magnitude, at every size and speed, so that the total magnetic contribution $\Delta L_H=\Delta L_H^{\rm int}+\Delta L_H^{\rm scat}$ is negative throughout, reinforcing the electric one, in agreement with Ref.~\cite{castellanos2023theory} at $a=5$~nm and extending it to $a=50$~nm. Unlike for aluminum, however, $|\Delta L_H^{\rm int}|$ grows with $v$ at every size, by a factor of $\approx2.5$ between $v=0.5c$ and $v=0.95c$ at $a=50$~nm, so that for gold the magnetic channel gains weight both through the decay of the electric terms and through its own growth. The direct magnitude comparison between gold and aluminum at fixed $(v,b)$ is addressed separately below, for the frequency-integrated total at $a=50$~nm, by Fig.~\ref{fig:Al_vs_Au}.

\begin{figure*}
    \centering
    \includegraphics[width=0.76\linewidth]{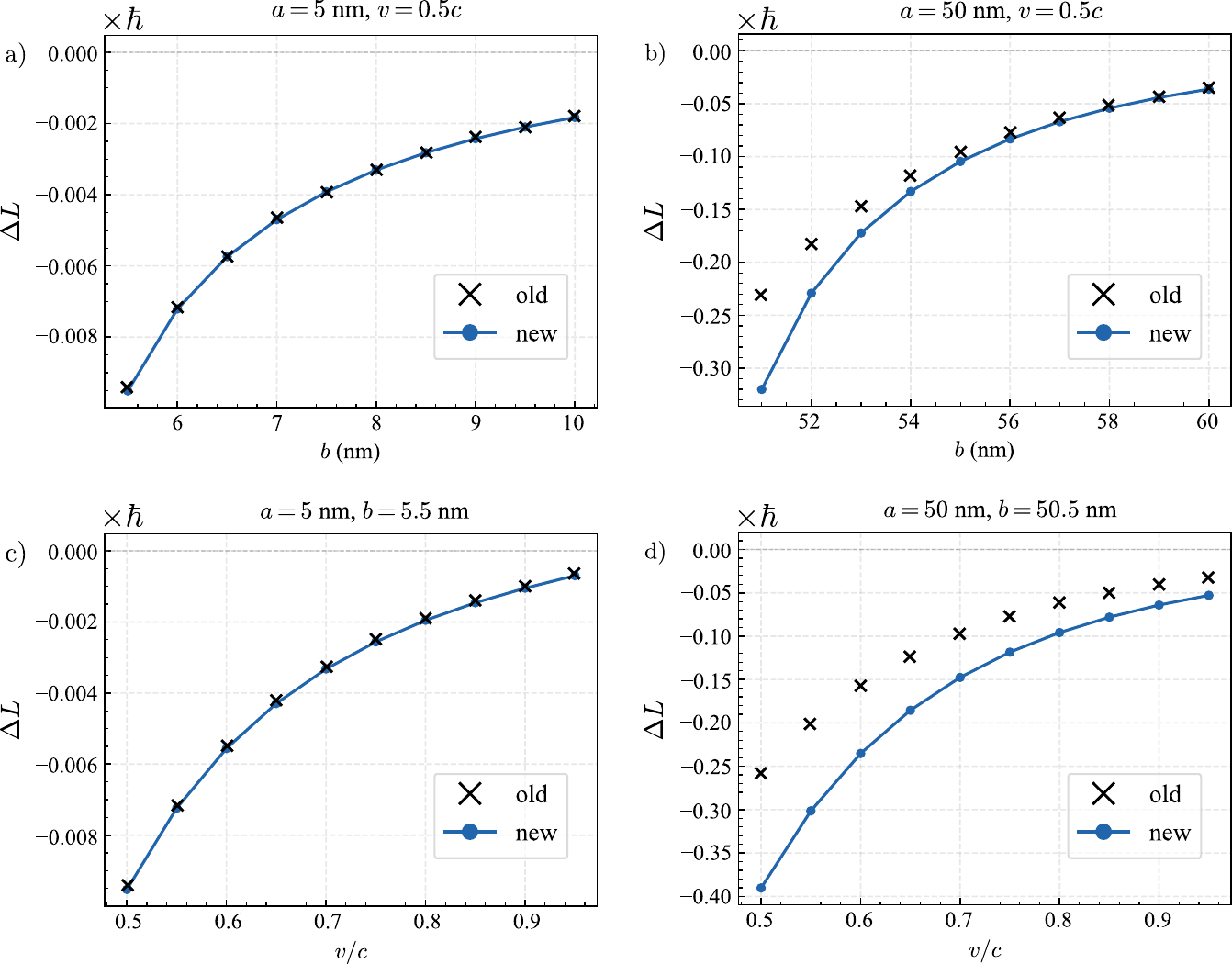}
    \caption{Benchmark of the present methodology against the digitized results of Ref.~\cite{castellanos2023theory} for a Drude aluminum nanoparticle, comparing their reported multipole truncation ($\ell_{\max}=10$ for $a=5$~nm, $\ell_{\max}=13$ for $a=50$~nm; crosses) against the present, fully converged calculation ($\ell_{\max}=15$-$51$; circles and solid lines). a) $a=5$~nm, $\Delta L$ vs.\ impact parameter $b$ at fixed $v=0.5c$ [their Fig.~3(a)]. b) $a=50$~nm, same trajectory family [their Fig.~3(g)]. c) $a=5$~nm, $\Delta L$ vs.\ $v$ at fixed $b=5.5$~nm [their Fig.~3(b)]. d) $a=50$~nm, $\Delta L$ vs.\ $v$ at fixed $b=50.5$~nm [their Fig.~3(h)].}
    \label{fig:benchmark_castellanos}
\end{figure*}
%%%%%%%%%%%%%%%%%%%%%%%%%%%%%%%%%%%%%%%%%%%%%%%%%%%%%%%%%%%%%%%%%
\subsection{Comparison between materials, and with previous large-particle calculations}\label{sec:comparison}
%%%%%%%%%%%%%%%%%%%%%%%%%%%%%%%%%%%%%%%%%%%%%%%%%%%%%%%%%%%%%%%%%

Figure~\ref{fig:Al_vs_Au} directly compares aluminum and gold at the electron-tweezer-relevant size $a=50$~nm, mapping $\Delta L$ over the full $(v,b)$ plane studied. Both materials show the same qualitative dependence, $|\Delta L|$ maximal at low speed and small impact parameter, decaying monotonically as either $v$ or $b$ increases, except for the speed dependence of gold at large $b$ (see below), and the same sign throughout (counterclockwise in Fig.~\ref{fig:system}), confirming at the full electron-tweezer-relevant size the qualitative picture established in the small-particle regime~\cite{castellanos2023theory}. Quantitatively, however, gold transfers substantially more angular momentum than aluminum across most of the $(v,b)$ plane explored: at the most favorable trajectory sampled, $v=0.5c$ and $b\to50$~nm (close to the surface), $\Delta L\approx-0.9 \hbar$ for gold versus $\Delta L\approx-0.39 \hbar$ for aluminum, a factor of $\sim\!2.3$. We attribute this to the richer interband structure of gold, which provides spectral weight across a much wider band of the near-field spectrum of the electron than the narrow plasmonic band of aluminum [Fig.~\ref{fig:spectral_density}(b) vs.\ Fig.~\ref{fig:maps_Au30}(a)], more than compensating for the lower plasma frequency of gold in setting the overall transferred angular momentum. This broader frequency range can be quantified directly: converging $\Delta L$ to within $1\%$ requires integrating out to $\hbar\omega_{\rm cut}\approx107$~eV for gold, roughly nine times the $\hbar\omega_{\rm cut}\approx12.5$~eV sufficient for aluminum (Fig.~\ref{fig:freq_cutoff}, Appendix~\ref{app: multipole convergence}), a direct, quantitative measure of how much more of the frequency domain contributes to the angular momentum transferred to gold. This picture is also consistent with the two departures from the behavior described above, both found at large impact parameter, where the near field of the electron reaches the nanoparticle only over a narrower frequency range that widens with increasing $v$: for $v\lesssim0.6c$ and $b\gtrsim55$~nm, aluminum transfers more angular momentum than gold, by up to $\approx30\%$, and for gold at $b\gtrsim53$~nm, $|\Delta L|$ no longer decreases monotonically with $v$, passing instead through a shallow minimum at high speed and becoming nearly independent of $v$ for $b\gtrsim59$~nm, where it varies by less than $10\%$ over the entire speed range. This direct, equal-footing comparison, same size, same trajectories, same multipole order, is made possible by the $\ell_{\max}=51$ convergence established in Appendix~\ref{app: multipole convergence}, and was previously inaccessible for gold at this particle size and multipole order.

A complementary, and more directly diagnostic, benchmark is available against Ref.~\cite{castellanos2023theory}, which reports $\Delta L$ for a Drude aluminum nanoparticle at $\ell_{\max}=10$ for $a=5$~nm and $\ell_{\max}=13$ for $a=50$~nm [their Fig.~3(a)-(b) and 3(g)-(h), respectively]. We digitized these four curves and compare them in Fig.~\ref{fig:benchmark_castellanos} against the present, fully converged calculation at the same size, material, and trajectories. At $a=5$~nm, where $\ell_{\max}=10$ already sits at the edge of the $\pm1\%$ convergence band established independently in Fig.~\ref{fig:Drude_convergence} (Appendix~\ref{app: multipole convergence}), the two calculations agree closely: the mean (maximum) relative deviation from the present calculation is $0.8\%$ ($2.1\%$) for the impact-parameter scan [Fig.~\ref{fig:benchmark_castellanos}a)] and $3.2\%$ ($8.6\%$) for the speed scan [Fig.~\ref{fig:benchmark_castellanos}c)]; in the latter the relative deviation grows toward the highest speeds, where $|\Delta L|$ is smallest and the magnetic contribution largest [Fig.~\ref{fig:DL_v_Al}(a)], while the absolute difference stays below $1.5\%$ of the range of $\Delta L$ spanned by the scan. Because the dipolar term ($\ell_{\max}=1$) alone already captures most of the converged magnitude of $\Delta L$ at this size (Fig.~\ref{fig:Drude_convergence}a, Appendix~\ref{app: multipole convergence}), this $a=5$~nm agreement also functions as an indirect validation of the dipolar (small-particle) limit of the present fully retarded formalism, complementing the direct analytic dipole-limit benchmark already reported by Ref.~\cite{castellanos2023theory} at $a=1$~nm. At $a=50$~nm, where $\ell_{\max}=13$ is nearly four times smaller than the $\ell_{\max}=51$ used here, the deviation grows substantially, to $10.7\%$ ($27.9\%$) and $35.0\%$ ($38.8\%$) for the same two trajectory families [Figs.~\ref{fig:benchmark_castellanos}b) and~\ref{fig:benchmark_castellanos}d)], respectively, with the truncated calculation systematically underestimating $|\Delta L|$: in the impact-parameter scan the deviation is concentrated at the smallest $b$ sampled, precisely where the electron-nanoparticle coupling, and with it the required multipole order, is largest, while for trajectories passing close to the surface ($b=50.5$~nm) it amounts to a roughly constant fraction, $33$--$39\%$, of the converged $\Delta L$ across the entire speed range.

This pattern, close agreement at $a=5$~nm and systematic disagreement at $a=50$~nm that grows toward the strongest near-field coupling (smallest $b$), is exactly what is expected if $\ell_{\max}=13$ is not yet converged at the larger size: an independently sourced, previously published data set now provides direct quantitative support for the convergence argument of Appendix~\ref{app: multipole convergence}, rather than only the single-size, extrapolated argument available previously. We do not attempt a matched-$\ell_{\max}=10$ or $13$ rerun of the present code to isolate this effect from possible differences between the two numerical implementations (adaptive cubature versus the present analytical angular integration), since doing so would defeat the purpose of the method introduced here: the analytical reduction of Sec.~\ref{th} is designed precisely to make $\ell_{\max}=13$ unnecessary, reaching $\ell_{\max}=51$ at $a=50$~nm at a fraction of the computational cost (Appendix~\ref{app: multipole convergence}). The convergence analysis of Appendix~\ref{app: multipole convergence} already establishes, independently of this benchmark, that $a=50$~nm requires $\ell_{\max}$ well beyond $13$ for quantitative convergence; the agreement/disagreement pattern reported here is consistent with, and adds independent support to, that conclusion, rather than resting on it alone.

To translate this frequency-integrated result into an experimentally relatable scale, consider a $100$~pA STEM probe ($\approx6.2\times10^{8}$ electrons/s) delivering the peak transferred angular momentum found here, $|\Delta L|\approx0.9\hbar$ per electron, to an isolated $a=50$~nm gold sphere of moment of inertia $I=\tfrac{2}{5}Ma^2\approx1.0\times10^{-32}$~kg\,m$^2$ (bulk gold density): the resulting torque, $\tau=(6.2\times10^{8})|\Delta L|\approx5.9\times10^{-26}$~N\,m, would produce, in the complete absence of damping, an angular acceleration $\alpha=\tau/I\approx5.9\times10^{6}$~rad/s$^2$, sufficient to spin the nanoparticle up to its room-temperature ($T=300$~K) rotational thermal scale, $\omega_{\rm th}=\sqrt{k_BT/I}\approx6.4\times10^{5}$~rad/s, within $\sim0.1$~s of continuous irradiation. This idealized, undamped estimate neglects rotational friction, substrate coupling, and beam-induced heating, all of which would raise the current or exposure time required in practice, but it indicates that the torques computed here fall within a range that could plausibly drive nanoparticle rotation at experimentally resolvable rates under realistic STEM operating conditions.

%*********************************************************%
\section{Conclusions}
\label{sec:conclusions}
%*********************************************************%

We have presented a fully causal, multipole-converged electrodynamical methodology to compute the angular momentum transferred from a swift electron to an isolated spherical nanoparticle, and an efficient numerical implementation that evaluates its spectral density through an analytic reduction to irreducible angular integrals rather than through numerical angular cubature. This reduction lowers the effective cost of the double multipolar sum from $O(\ell_{\max}^4)$ to $O(\ell_{\max}^3)$ and allowed us to extend the multipole truncation to $\ell_{\max}=51$ for nanoparticles with radius as large as $a=50$~nm, which is nearly four times the order reported at this same particle size in Ref.~\cite{castellanos2023theory}, simultaneously for a Drude metal and for a material with a substantially more complex, interband-dominated dielectric response, and, for gold specifically, at a particle size ten times larger than previously reached for an optically complex material. This approach ensures controlled numerical accuracy and allows systematic analysis of the physical mechanisms governing angular momentum transfer across the nanoscale, at particle sizes directly relevant to electron-tweezer experiments and previously accessible only through frequency-integrated, single-material calculations.

Using causal dielectric functions for aluminum and gold, we have analyzed the dependence of the angular momentum transfer on electron speed, impact parameter, nanoparticle size, and material-specific resonances. Aluminum, described by a Drude response, exhibits a single cluster of resonances between $5$ and $9$~eV that splits into several closely spaced multipolar peaks as the nanoparticle radius grows from $5$ to $30$~nm. Gold, whose interband transitions are superimposed on a plasmonic free-electron response, displays a correspondingly broader and richer spectral structure extending continuously from about $5$ to $40$~eV. Consistent with the frequency-integrated results of Ref.~\cite{castellanos2023theory}, we find that the net angular momentum transferred to both materials remains of the same sign (counterclockwise relative to the $\hat{y}$ direction in Fig.~\ref{fig:system}) throughout the parameter range considered; here we further show that this conclusion, previously established only from the frequency-integrated response and only up to $\ell_{\max}=13$ at $a=50$~nm, persists once the full spectral density is resolved and the multipole expansion is converged several times further. At equal size and trajectory, gold transfers substantially more angular momentum than aluminum at $a=50$~nm, by a strongly speed-dependent factor ranging from about $2\times$ at $v=0.5c$ to about $5\times$ at $v=0.95c$ (fixed $b=51$~nm), showing that the strength of the effect is set not simply by a plasma frequency of the material but by the full structure of its causal response across the near-field bandwidth of the electron.

Resolving the spectral density also lets us trace this behavior to its physical origin across the full frequency domain, rather than only in the frequency-integrated total. We confirm and extend the finding of Ref.~\cite{castellanos2023theory} that the angular momentum transfer is dominated by the interaction term of the Maxwell stress tensor, the interference between the external field of the swift electron and the field scattered by the nanoparticle, while the scattered-scattered contribution, though non-negligible and of opposite sign, remains subdominant in magnitude at essentially every frequency, not merely in its integral. Moreover, resolving $\mathcal{L}(\omega)$ shows that this interference-dominated structure holds separately in the electric and magnetic channels, with the magnetic-interaction spectral density closely tracking the resonances of the electric one at a level between a factor of $\approx4$ and $\approx20$ below it for aluminum, depending on size and frequency, and a nearly constant factor of $\approx5$-$7$ below it for gold. This indicates that angular momentum transfer in electron-nanoparticle interactions is fundamentally a near-field, interference-driven phenomenon, in which the magnetic field of the swift electron plays a subdominant role that is nonetheless non-negligible for certain parameters.

For both materials, the electric contribution dominates the magnetic one at low speed at every size from $5$ to $50$~nm, consistent with Ref.~\cite{castellanos2023theory} at $a=5$~nm and extending it, for gold, up to $a=50$~nm, which was not reported previously. The magnetic contribution, however, carries the same sign as the electric one at every size and speed examined, in agreement with Ref.~\cite{castellanos2023theory}, and its relative weight grows markedly with electron speed, from a few percent of $\Delta L$ at $v=0.5c$ to between a quarter and a half of it at $v=0.95c$ for trajectories passing close to the surface, becoming comparable to the electric contribution for aluminum at $a=50$~nm.

Within the assumptions of the present model, namely isolated spherical nanoparticles in vacuum described by a local dielectric response, our results demonstrate that a reversal in the direction of the angular momentum transfer does not occur. Consequently, experimentally observed direction-selective behavior must originate from physical mechanisms not captured by the present model. These may include substrate-mediated forces, nanoparticle charging and secondary-electron emission, nonlocal and quantum-size effects in the dielectric response, thermal gradients, or deviations from spherical geometry.

By establishing a clear, causal, and spectrally resolved baseline for angular momentum transfer at nanoparticle sizes relevant to electron-tweezer applications, this work delineates the limits of applicability of isolated-particle electrodynamic models and provides quantitative benchmarks for future theoretical and experimental studies. The analytical and computational methodology introduced here provides a starting point for investigating more complex scenarios, including nonspherical, chiral, or magnetic nanoparticles, collective interactions in nanoparticle assemblies, and the simultaneous transfer of linear and angular momentum. Such extensions will be essential for developing predictive models of electron-beam-based nanoscale manipulation and for advancing the concept of electron tweezers beyond idealized systems.
%%%%%%%%%%%%%%%%%%%%%%%%%%%%%%%%%%%%%%%%%%%%%%%%%%%
\begin{acknowledgments}
This work was supported by the UNAM-PAPIIT DGAPA IN101825 project. This research was performed using services/resources provided by Grid UNAM, which is a collaborative effort driven by DGTIC and the research Institutes of Astronomy, Nuclear Sciences, and Atmosphere Sciences and Climate Change at UNAM, within the framework of the TMERN project. J. L. B-G. acknowledges a PhD scholarship from Secretar\'ia de Ciencia, Humanidades, Tecnolog\'ia e Innovaci\'on (Secihti), Mexico.
\end{acknowledgments}

%**************************************************
%************* APENDIX ***************************
%************************************************
%%%%%%%%%%%%%%%%%%%%%%%%%%%%%%%%%%%%%%%%%%%
\appendix

%%%%%%%%%%%%%%%%%%%%%%%%%%%%%%%%%%%%%%%%%%%%%%%%%%%%%%%%%%%%%%%%%%%
\section{Numerical convergence and computational cost}\label{app: multipole convergence}
%%%%%%%%%%%%%%%%%%%%%%%%%%%%%%%%%%%%%%%%%%%%%%%%%%%%%%%%%%%%%%%%%%%%

All results reported in the main text are obtained from the fully retarded multipole expansions of the external and scattered fields in Eqs.~\eqref{eq:ScatterdEField} and~\eqref{eq:ScatterdMField}, together with the Maxwell stress tensor expression for the spectral density in Eq.~\eqref{eq: spectral density AMT}. In practice, these formulas involve (i) infinite sums over multipole order $\ell$ and azimuthal index $m$ in the field expansions, and (ii) corresponding double sums over $(\ell,m)$ and $(\ell',m')$ when forming the quadratic field products that appear in the stress tensor, e.g., Eq.~\eqref{Tradial} and Eq.~\eqref{eq: projection Maxwell Tensor}. Numerical evaluation therefore requires truncating the multipole sums at a maximum order $\ell_{\max}$, i.e., $\sum_{\ell=1}^{\infty}\rightarrow\sum_{\ell=1}^{\ell_{\max}}$ (and analogously for $\ell'$), and, separately, truncating the frequency integral that yields $\Delta L_n$ (Sec.~\ref{th}) at a finite upper cutoff $\omega_{\rm cut}$. The purpose of this Appendix is to document the convergence of the angular-momentum transfer with respect to both truncations, and its consequence for computational cost.

%%%%%%%%%%%%%%%%%%%%%%%%%%%%%%%%%%%%%%%%%%%%%%%%%%%%%%%%%%%%%%%%%%%%
\subsection{Multipole truncation}
%%%%%%%%%%%%%%%%%%%%%%%%%%%%%%%%%%%%%%%%%%%%%%%%%%%%%%%%%%%%%%%%%%%%

\begin{figure}
    \centering
    \includegraphics[width=0.9\linewidth]{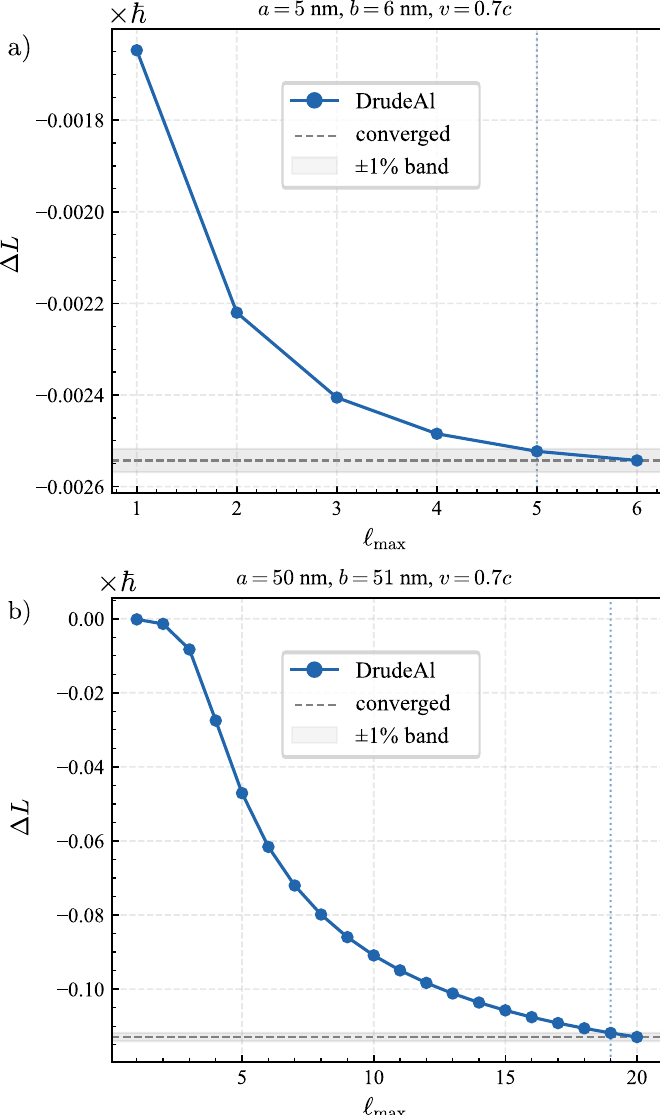}
    \caption{Convergence of the frequency-integrated angular momentum transfer $\Delta L$ with respect to the multipole truncation $\ell_{\max}$, for a Drude aluminum nanoparticle at the two extremes of particle size studied in this work. a) $a=5$~nm, $b=6$~nm, $v=0.7c$; value at the last order computed, $\ell_{\max}=6$, where the change between successive orders first falls below $1\%$: $\Delta L\approx-2.5\times10^{-3}\,\hbar$, reached to within $\pm1\%$ by $\ell_{\max}\approx5$. b) $a=50$~nm, $b=51$~nm, $v=0.7c$; last order computed $\ell_{\max}=20$: $\Delta L\approx-1.1\times10^{-1}\,\hbar$, reached to within $\pm1\%$ by $\ell_{\max}\approx19$. Dashed lines and shaded bands mark the value at the last order computed and $\pm1\%$ about it in each panel; dotted vertical lines mark the $\ell_{\max}$ at which the curve enters and remains within this band.}
    \label{fig:Drude_convergence}
\end{figure}

To this end, we consider two representative test cases spanning the particle-size range studied in the main text, both for a Drude aluminum nanoparticle in vacuum (see Fig.~\ref{fig:system}), using the same dielectric parameters as in the main text ($\hbar\omega_p=13.14$~eV and $\hbar\Gamma=0.197$~eV~\cite{Markovic}): $a=5$~nm at $b=6$~nm and $a=50$~nm at $b=51$~nm, both at $v=0.7c$ and matching the trajectory family of Fig.~\ref{fig:DL_v_Al} ($b=a+1$~nm), so that the two panels differ only in particle size and isolate its effect on multipole convergence. For each case, $\ell_{\max}$ is increased from $1$ until the relative change of $\Delta L$ between successive orders falls below $1\%$; the figure marks the order from which the curve remains within $\pm1\%$ of the last value computed. This threshold is used here, and in Fig.~\ref{fig:cost_scaling} below, purely for visual clarity: a band this wide is legible on the scale of Fig.~\ref{fig:Drude_convergence}, whereas the production runs use a threshold of $10^{-4}$, on the change between successive orders for $\ell_{\max}$ (up to $\ell_{\max}=51$, Sec.~\ref{results}) and on the neglected tail for the frequency cutoff of Fig.~\ref{fig:freq_cutoff} in the next subsection, a band that would be visually indistinguishable from the converged value at the scale shown here.

Figure~\ref{fig:Drude_convergence} shows $\Delta L$ as a function of $\ell_{\max}$ for both test cases. In both panels the convergence is monotonic in magnitude, with no sign change at any truncation order. The two sizes differ sharply, however, in where this growth sets in. At $a=5$~nm [panel a)], the dipolar term ($\ell_{\max}=1$) alone already captures most of the converged magnitude, and the curve enters the $\pm1\%$ band by $\ell_{\max}\approx5$. At $a=50$~nm [panel b)], by contrast, the dipolar and quadrupolar terms ($\ell_{\max}=1,2$) contribute almost nothing: $\Delta L$ is indistinguishable from zero at this scale through $\ell_{\max}=2$, and only grows appreciably once several more multipolar orders are included, entering the $\pm1\%$ band only by $\ell_{\max}\approx19$. Because the successive changes decay slowly for trajectories passing close to the surface, stopping at a $1\%$ change still leaves a remainder: the production value at $\ell_{\max}=51$, $\Delta L\approx-0.123\,\hbar$ (Fig.~\ref{fig:freq_cutoff}), lies $9\%$ beyond the value at $\ell_{\max}=20$, which is why the production runs use the far stricter criterion of Sec.~\ref{results}. A dipolar or small-particle truncation at this size would therefore miss not merely the correct magnitude of the transferred torque but essentially the entire effect, underscoring why a converged multipolar treatment is indispensable at electron-tweezer-relevant sizes.

This growth in required multipole order with particle size, and its cost, can be quantified directly. Figure~\ref{fig:cost_scaling}(a) shows the $1\%$-converged $\ell_{\max}$ against particle radius for the full set of sizes studied, $a=1$-$50$~nm, at a fixed trajectory $b=a+1$~nm ($b=1.5$~nm for $a=1$~nm), $v=0.7c$: $\ell_{\max}$ grows smoothly from $4$ at $a=1$~nm to $20$ at $a=50$~nm. The production runs reported in Sec.~\ref{results} use $\ell_{\max}=44$ for aluminum and $\ell_{\max}=51$ for gold at $a=30$~nm, and $\ell_{\max}=51$ for both materials at $a=50$~nm. These values exceed the $1\%$-converged $\ell_{\max}$ shown in panel (a) because production runs are pushed to a relative change of $10^{-4}$ between successive orders, or to $\ell_{\max}=51$ (Sec.~\ref{results}), rather than to the $1\%$ threshold used for illustration in this figure; each value is, in every case, independently verified for the specific material, frequency, and trajectory considered. Each of the $a=30$ and $a=50$~nm values substantially exceeds the $\ell_{\max}=13$ reported at $a=50$~nm in Ref.~\cite{castellanos2023theory}.

Figure~\ref{fig:cost_scaling}(b) tests directly the $O(\ell_{\max}^3)$ scaling argued for in Sec.~\ref{th C}: plotting the wall-clock time to reach the converged $\ell_{\max}$ at each size against that $\ell_{\max}$ on logarithmic axes gives an empirical power law $t\propto\ell_{\max}^{2.7}$, clearly inconsistent with the naive $O(\ell_{\max}^4)$ scaling of an unreduced double multipolar sum and, if anything, mildly more favorable than the target $O(\ell_{\max}^3)$. We attribute this mild undershoot chiefly to a roughly $\ell_{\max}$-independent overhead (solver initialization, dielectric-function evaluation, and file I/O) that is proportionally largest at the smallest $\ell_{\max}$ sampled and biases a single power-law fit across the full range toward a shallower apparent slope than the asymptotic, large-$\ell_{\max}$ behavior. Either way, the measured scaling falls well short of $O(\ell_{\max}^4)$, providing direct empirical support, beyond the analytical argument of Sec.~\ref{th C}, that the azimuthal selection rule delivers a real computational advantage across the full range of particle sizes studied here.
\begin{figure}
    \centering
    \includegraphics[width=0.95\linewidth]{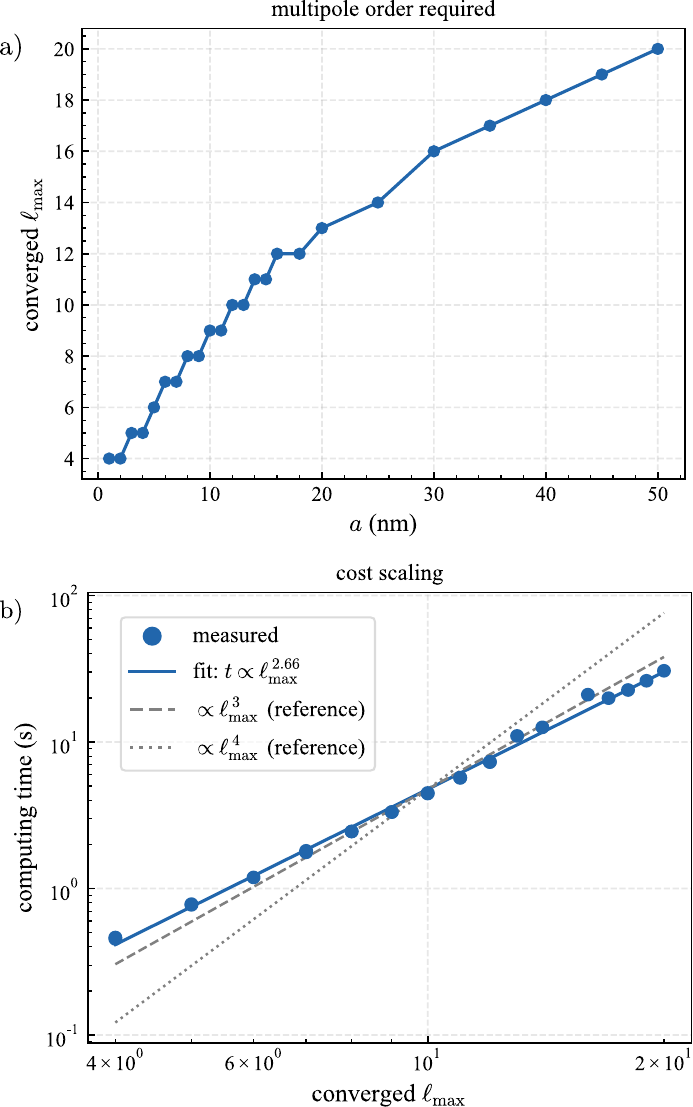}
    \caption{Empirical test of the $O(\ell_{\max}^3)$ computational-scaling argument of Sec.~\ref{th C}, for a Drude aluminum nanoparticle at the fixed trajectory $b=a+1$~nm ($b=1.5$~nm for $a=1$~nm), $v=0.7c$, across the full range of particle sizes studied in this work ($a=1$-$50$~nm). a) Multipole order $\ell_{\max}$ at which the relative change of $\Delta L$ between successive orders first falls below $1\%$, as a function of particle radius $a$. b) Wall-clock computing time to reach that converged $\ell_{\max}$, plotted against $\ell_{\max}$ on logarithmic axes. The solid line is a least-squares power-law fit to the full data set, $t\propto\ell_{\max}^{2.7}$; dashed and dotted gray lines show $\ell_{\max}^3$ and $\ell_{\max}^4$ references, respectively, anchored to the fit at the median $\ell_{\max}$ sampled for visual comparison of slope only.}
    \label{fig:cost_scaling}
\end{figure}

For context, Ref.~\cite{castellanos2021phdthesis} reports wall-clock computing times for this class of calculation obtained via adaptive numerical cubature of the angular integrals, the strategy replaced by the analytical reduction of Sec.~\ref{th C}: for a Drude aluminum nanoparticle at $v=0.5c$, reaching $\ell_{\max}=13$ at $a=10$~nm required $\approx21.5$~h, and reaching $\ell_{\max}=16$ at $a=50$~nm required $\approx19.7$~h, on a laptop-class machine~\footnote{Intel Core i5-10400H at $2.60$~GHz, per Ref.~\cite{castellanos2021phdthesis}.}. The corresponding times read from Fig.~\ref{fig:cost_scaling}(b) are $\approx11$~s and $\approx21$~s, respectively, measured on a modest desktop machine~\footnote{Intel Core i3-8100 at $3.60$~GHz. Its PassMark single-thread rating (2186) is, if anything, $15\%$ \emph{lower} than that reported for the laptop above (2567; PassMark Software, \url{https://www.cpubenchmark.net}, accessed August 2026), so hardware differences cannot account for any part of the gap reported here, and if anything work against it.}. Even allowing for the imprecision of reading both figures by eye, the reduction in computing time exceeds three orders of magnitude (roughly $3\times10^3$-$7\times10^3$, depending on the size compared), consistent with the combined effect of the improved asymptotic scaling established above and the replacement of adaptive angular cubature by machine-precision analytical quadrature.

%%%%%%%%%%%%%%%%%%%%%%%%%%%%%%%%%%%%%%%%%%%%%%%%%%%%%%%%%%%%%%%%%%%%
\subsection{Frequency-domain truncation}
%%%%%%%%%%%%%%%%%%%%%%%%%%%%%%%%%%%%%%%%%%%%%%%%%%%%%%%%%%%%%%%%%%%%

The frequency integral defining $\Delta L_n$ (Sec.~\ref{th}) extends formally to $\omega\to\infty$ and is evaluated numerically, via Gauss-Kronrod quadrature~\cite{castrejon2021effects}, up to a finite cutoff $\omega_{\rm cut}$. To select $\omega_{\rm cut}$ for each production run, the tail contribution $\int_{\omega_{\rm cut}}^\infty\mathcal{L}_n(\omega)\,d\omega$ was evaluated independently using double-exponential (tanh-sinh) quadrature \cite{takahasi1974double}, well suited to this semi-infinite, rapidly decaying integrand, and $\omega_{\rm cut}$ was increased until this tail integral fell below the $10^{-4}$ relative-error threshold used throughout this work.

\begin{figure}
    \centering
    \includegraphics[width=1.0\linewidth]{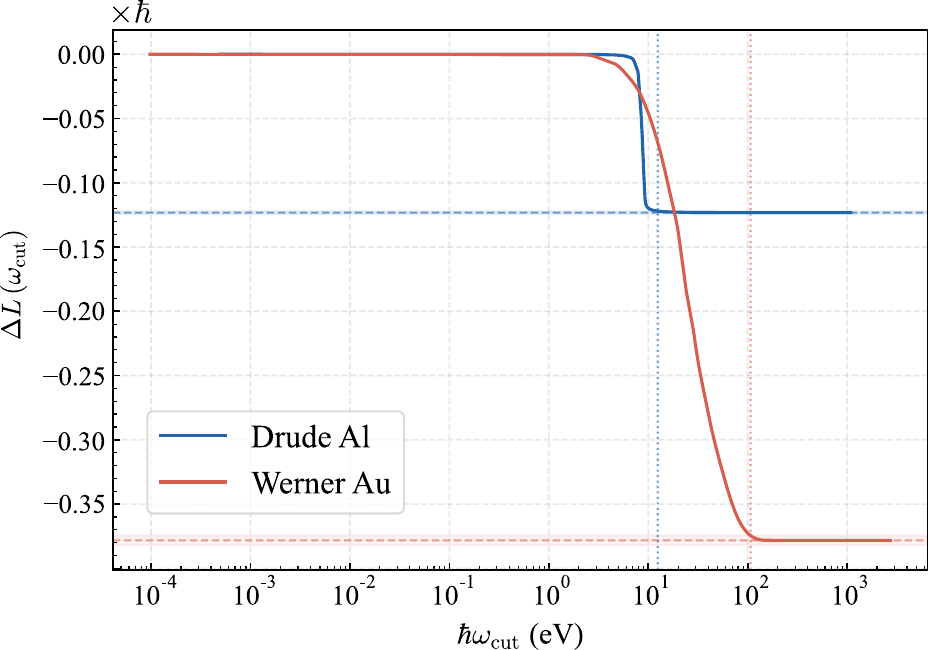}
    \caption{Convergence of the frequency-integrated angular momentum transfer $\Delta L$ with respect to the frequency-integration cutoff $\omega_{\rm cut}$, for Drude aluminum (blue) and Werner gold (red) nanoparticles at $a=50$~nm, $b=51$~nm, $v=0.7c$. Dashed lines and shaded bands mark the converged value and $\pm1\%$ about it for each material; dotted vertical lines mark the $\hbar\omega_{\rm cut}$ at which each curve enters and remains within this band.}
    \label{fig:freq_cutoff}
\end{figure}

Figure~\ref{fig:freq_cutoff} visualizes this convergence directly, independently of that analytical bound, for the $a=50$~nm, $b=51$~nm, $v=0.7c$ trajectory of Fig.~\ref{fig:DL_v_Al}c) and its gold analog, by plotting the cumulative integral $\Delta L(\omega_{\rm cut})=\int_0^{\omega_{\rm cut}}\mathcal{L}(\omega)\,d(\hbar\omega)$ against $\omega_{\rm cut}$ itself, for both materials on a common logarithmic frequency axis.
Both curves are flat, at $\Delta L=0$, for $\hbar\omega_{\rm cut}$ below the onset of the lowest resonance, rise through the resonance region identified in Fig.~\ref{fig:spectral_density}, and then plateau: Drude aluminum enters and remains within the $\pm1\%$ band of its converged value, $\Delta L\approx-0.123\,\hbar$, by $\hbar\omega_{\rm cut}\approx12.5$~eV, while gold, consistent with the broader interband-driven plateau of its spectral density (Fig.~\ref{fig:maps_Au30}a), requires roughly nine times more, entering the band only by $\hbar\omega_{\rm cut}\approx107$~eV and converging to $\Delta L\approx-0.378\,\hbar$. As in Fig.~\ref{fig:Drude_convergence}, the $\pm1\%$ band plotted here is a visual reference only, chosen to be legible at the scale of this figure; the frequency cutoffs actually used in every production run, including this one, satisfy the tighter $10^{-4}$ relative-error threshold defined above. Both curves remain flat well beyond these points, out to the largest cutoffs tested ($\gtrsim1$~keV for aluminum, $\gtrsim2.7$~keV for gold), with no further drift: direct visual confirmation, for a representative case, that the production cutoffs used throughout this work leave no unconverged tail contribution, complementing the analytical bound described above.

%%%%%%%%%%%%%%%%%%%%%%%%%%%%%%%%%%%%%%%%%%%%
\section{Multipole expansion of the external and the scattered electromagnetic fields} \label{app: multipole expansion}

For completeness, we reproduce here the multipole expansion of the external and scattered electromagnetic fields, identical to that of Ref.~\cite{castrejon2026electrodynamics} and originally derived in Ref.~\cite{de1999relativistic}, that forms the starting point for the analytical evaluation of the Maxwell stress tensor and the angular momentum transfer computed in Appendix~\ref{app: momentum spectral density}. Rather than leaving it implicit in Refs.~\cite{castrejon2021time,castellanos2023theory}, we make explicit the Fourier convention and the frequency-folding step that together produce the factor of $1/\pi$ in Eq.~\eqref{eq: spectral density AMT}. Each real-valued time-domain field component is defined through
\begin{equation}
f(\vb{r},t)=\int_{-\infty}^{\infty}\frac{d\omega}{2\pi}\,f(\vb{r};\omega)\,e^{-i\omega t},
\label{eq: FT convention}
\end{equation}
so that reality of $f(\vb{r},t)$ enforces $f(\vb{r};-\omega)=f^{*}(\vb{r};\omega)$. The physical angular momentum transfer is the time integral $\Delta L_i=\int_{-\infty}^{\infty}\!dt\oint_S \epsilon_{i}^{\,\,\,\,lj}r_l\,\mathscr{T}_{jk}(\vb{r},t)\,dS^k$ of the real, time-domain stress tensor, which is quadratic in the fields. Parseval's theorem converts each such quadratic product into a two-sided frequency integral, e.g., $\int dt\,E_j(\vb{r},t)E_k(\vb{r},t)=\int_{-\infty}^{\infty}\frac{d\omega}{2\pi}E_j(\vb{r};\omega)E_k^{*}(\vb{r};\omega)$; the reality condition then folds its $\omega<0$ half onto $\omega>0$, since $E_j(\vb{r};-\omega)E_k^{*}(\vb{r};-\omega)=[E_j(\vb{r};\omega)E_k^{*}(\vb{r};\omega)]^{*}$, giving $\int_{-\infty}^{\infty}\frac{d\omega}{2\pi}(\cdots)=\int_{0}^{\infty}\frac{d\omega}{\pi}\Re\{\cdots\}$. This single step simultaneously produces the explicit real part in the frequency-domain stress tensor of Eq.~\eqref{eq: maxwell tensor freq}, the $1/\pi$ prefactor in Eq.~\eqref{eq: spectral density AMT}, and the one-sided integration range $\omega\in[0,\infty)$ used throughout, guaranteeing that $\Delta L_n$ is manifestly real rather than only incidentally so.

The interaction between the electron and the NP is treated using a fully retarded wave solution, in which the total electromagnetic field is expressed as the superposition of the external field generated by the bare swift electron and the scattered field produced by the electric charge and current densities induced within the NP. Both external and scattered electromagnetic fields are expanded in spherical coordinates as multipole series in the frequency domain \cite{de1999relativistic, castrejon2021time}:
\begin{align}
    \vb{E}^{\text{(e,s)}}(\vb{r};\omega)= \sum_{\ell=1}^{\infty}\sum_{m=-\ell}^{\ell} \Big[&
    \scr{E}_{\ell,m}^{\text{(e,s)} r}\,\hb{r} +
    \scr{E}_{\ell,m}^{\text{(e,s)} \theta}\,\hat{\boldsymbol{\theta}} \nonumber \\
    &+
    \scr{E}_{\ell,m}^{\text{(e,s)} \varphi}\,\hat{\boldsymbol{\varphi}}\Big], \label{eq:ScatterdEField} \\
    \vb{H}^{\text{(e,s)}}(\vb{r};\omega)=  \sum_{\ell=1}^{\infty}\sum_{m=-\ell}^{\ell} \Big[&
    \scr{H}_{\ell,m}^{\text{(e,s)} r}\,\hb{r}+
    \scr{H}_{\ell,m}^{\text{(e,s)} \theta}\,\hat{\boldsymbol{\theta}} \nonumber \\
    &+\scr{H}_{\ell,m}^{\text{(e,s)} \varphi}\,\hat{\boldsymbol{\varphi}}\Big], \label{eq:ScatterdMField}
\end{align}
where the superscript $\text{(e,s)}$ denotes either the external ($\text{e}$) or scattered ($\text{s}$) field contribution. The explicit expressions for the multipole coefficients $\scr{E}_{\ell,m}^{\text{(e,s)}}$ and $\scr{H}_{\ell,m}^{\text{(e,s)}}$, expressed in SI units, are given below.

Electromagnetic fields are expressed in spherical coordinates $(r,\theta,\varphi)$, related to Cartesian coordinates through $x=r\sin\theta\cos\varphi$, $y=r\sin\theta\sin\varphi$, and $z=r\cos\theta$, with the $x,y$ and $z$ coordinates defined as shown in Fig.~\ref{fig:system} and the nanoparticle located at the origin.
\begin{align}
    \scr{E}_{\ell,m}^{\text{(e,s)}r}=&\,\rme ^{im\varphi}D_{\ell,m}^{\text{(e,s)}}\ell(\ell+1)P_\ell^m(\cos\theta)\frac{Z_\ell^{\text{(e,s)}}(kr)}{kr},
    \\
    \begin{split}
    \scr{E}_{\ell,m}^{\text{(e,s)}\theta}=&-\rme^{im\varphi}C_{\ell,m}^\text{(e,s)}\frac{m}{\sin\theta}Z_\ell^{\text{(e,s)}}(kr)P_\ell^m(\cos\theta)\\
    &-\rme^{im\varphi}D_{\ell,m}^\text{(e,s)}\left[(\ell+1)\frac{\cos\theta}{\sin\theta}P_\ell^m(\cos\theta)\right.\\
    &-\left.\frac{(\ell-m+1)}{\sin\theta}P_{\ell+1}^m(\cos\theta)\right]\\
    &\times\left[(\ell+1)\frac{Z_\ell^{\text{(e,s)}}(kr)}{kr}-Z_{\ell+1}^{\text{(e,s)}}(kr)\right],
    \end{split}
    \\
    \begin{split}
\scr{E}_{\ell,m}^{\text{(e,s)}\varphi}=&\,i\rme^{im\varphi}C_{\ell,m}^\text{(e,s)}Z_\ell^{\text{(e,s)}}(kr) \\
&\times \left[(\ell+1)\frac{\cos\theta}{\sin\theta}P_\ell^m(\cos\theta)\right.\\
    &\left.-\frac{(\ell-m+1)}{\sin\theta}P_{\ell+1}^m(\cos\theta)\right]\\
    &+i\rme^{im\varphi}D_{\ell,m}^\text{(e,s)}\frac{m}{\sin\theta}P_\ell^m(\cos\theta)\\
    &\times\left[(\ell+1)\frac{Z_\ell^{\text{(e,s)}}(kr)}{kr}-Z_{\ell+1}^{\text{(e,s)}}(kr)\right],
    \end{split}
\end{align}
and
\begin{align}
   \scr{H}_{\ell,m}^{\text{(e,s)}r}=&\,\rme^{im\varphi}C_{\ell,m}^\text{(e,s)}\ell(\ell+1)P_\ell^m(\cos\theta)\frac{Z_\ell^{\text{(e,s)}}(kr)}{kr},
    \\
    \begin{split}
    \scr{H}_{\ell,m}^{\text{(e,s)}\theta}=&\,\rme^{im\varphi}D_{\ell,m}^\text{(e,s)}\frac{m}{\sin\theta}Z_\ell^{\text{(e,s)}}(kr)P_\ell^m(\cos\theta)\\
    &-\rme^{im\varphi}C_{\ell,m}^\text{(e,s)}\left[(\ell+1)\frac{\cos\theta}{\sin\theta}P_\ell^m(\cos\theta)\right.\\
    &-\left.\frac{(\ell-m+1)}{\sin\theta}P_{\ell+1}^m(\cos\theta)\right]\\
    &\times\left[(\ell+1)\frac{Z_\ell^{\text{(e,s)}}(kr)}{kr}-Z_{\ell+1}^{\text{(e,s)}}(kr)\right],
    \end{split}
    \\
    \begin{split}
    \scr{H}_{\ell,m}^{\text{(e,s)}\varphi}=&\,i\rme^{im\varphi}C_{\ell,m}^\text{(e,s)}\frac{m}{\sin\theta}P_\ell^m(\cos\theta)\\
    &\times\left[(\ell+1)\frac{Z_\ell^{\text{(e,s)}}(kr)}{kr}-Z_{\ell+1}^{\text{(e,s)}}(kr)\right]\\
    &-\rme^{im\varphi}D_{\ell,m}^\text{(e,s)}Z_{\ell}^{\text{(e,s)}}(kr)  \\
    &\times \left[(\ell+1)\frac{\cos\theta}{\sin\theta}P_\ell^m(\cos\theta)\right.\\
    &-\left.\frac{(\ell-m+1)}{\sin\theta}P_{\ell+1}^m(\cos\theta)\right],
    \end{split}
\end{align}
where $P_\ell^m$ are the associated Legendre functions, and $Z_\ell^{(\mathrm{e})}=j_\ell(kr)$ ensuring regularity at the origin for the external-field expansion, and $Z_\ell^{(\mathrm{s})}=h_\ell^{(+)}(kr)$ enforcing the outgoing-wave condition for the scattered field \cite{castrejon2021time}, and $k$ is the vacuum wave number. The coefficients $C_{\ell,m}^{\text{(e,s)}}$ and $D_{\ell,m}^{\text{(e,s)}}$ are defined as
\begin{align}
    C_{\ell,m}^\text{(e,s)}&=(i)^\ell\sqrt{\frac{2\ell+1}{4\pi}\frac{(\ell-m)!}{(\ell+m)!}}\psi_{\ell,m}^\text{M,(e,s)},\\
    D_{\ell,m}^\text{(e,s)}&=(i)^\ell\sqrt{\frac{2\ell+1}{4\pi}\frac{(\ell-m)!}{(\ell+m)!}}\psi_{\ell,m}^\text{E,(e,s)},
\end{align}
where the explicit expressions for the coefficients $\psi_{\ell,m}^{\text{E,e}}$, $\psi_{\ell,m}^{\text{M,e}}$, $\psi_{\ell,m}^{\text{E,s}}$, and $\psi_{\ell,m}^{\text{M,s}}$ can be found in Ref.~\cite{de1999relativistic}.

%%%%%%%%%%%%%%%%%%%%%%%%%%%%%%%%%%%%%%%%%%%%%%%%%%%%%%%%%%%%%%%%%%%%%%%%%%%%%%%%%%%%%%%%%
\section{Expressions for the spectral density of the angular momentum transfer} \label{app: momentum spectral density}

In this Appendix, we provide the explicit expressions required to evaluate the spectral density of the angular momentum transfer defined in Eq.~\eqref{eq: spectral density AMT}. These expressions are obtained from the Maxwell stress tensor in the frequency-domain introduced in Eq.~\eqref{eq: maxwell tensor freq}, evaluated on a closed spherical integration surface that encloses the nanoparticle and does not intersect the electron trajectory.

Substituting the multipole expansions of Eq.~\eqref{eq:ScatterdEField} into the radial component of the electric contribution to the stress tensor, Eq.~\eqref{Tradial}, the products of the electromagnetic field components that appear there can be evaluated analytically. As an example, the product of the zebithal-radial electric-field components takes the form
\begin{align}
%
% Electric-Electric phir
%
\mathcal{E}_{
\ell, m}^{\text{(e,s)}\varphi}
\mathcal{E}_{\ell^{\prime}, m^{\prime}}^{(\text{e}^{\prime},\text{s}^{\prime})\text{r}\,*}
= \, &\Bigg\{C_{\ell, m}^{\rm \text{(e,s)}}D_{\ell^{\prime}, m^{\prime}}^{\rm (\text{e}^{\prime},\text{s}^{\prime}) \, *} \frac{Z_{\ell}^{\rm (e,s)}Z_{\ell^{\prime}}^{\rm (\text{e}^{\prime},\text{s}^{\prime}) \, *}}{k r} \nonumber\\
&\times \Bigg[(\ell +1) \frac{P_{\ell}^m P_{\ell^{\prime}}^{m^{\prime}} \cos\theta}{\sin\theta} \nonumber \\
& - (\ell -m + 1) \frac{P_{\ell + 1}^m P_{\ell^{\prime}}^{m^{\prime}}}{\sin\theta}\Bigg] \nonumber \\
& +D_{\ell, m}^{\rm \text{(e,s)}}D_{\ell^{\prime}, m^{\prime}}^{\rm (\text{e}^{\prime},\text{s}^{\prime}) \, *} m \frac{f_{\ell}^{\rm (e,s)}Z_{\ell^{\prime}}^{\rm (\text{e}^{\prime},\text{s}^{\prime}) \, *}}{k r} \nonumber \\
&\times \frac{P_{\ell}^m P_{\ell^{\prime}}^{m^{\prime}}}{\sin\theta}\Bigg\} i\, \ell^{\prime} (\ell^{\prime}+1)
\rme^{i (m-m^{\prime}) \varphi},
\label{eq: cenital radial electric-field stress component}
\end{align}
where $Z_\ell^{(\mathrm{e})}=j_\ell(kr)$ ensures regularity at the origin of the external-field expansion, and $Z_\ell^{(\mathrm{s})}=h_\ell^{(+)}(kr)$ enforces the outgoing-wave condition for the scattered field \cite{castrejon2021time}, and $f_{\ell}^{\rm (e,s)} = (\ell+1) Z_{\ell}^{\rm (\text{e},\text{s})}/(k r) -Z_{\ell+1}^{\rm (\text{e},\text{s})}$. The coefficients $C_{\ell,m}^{(\text{e,s})}$ and $D_{\ell,m}^{(\text{e,s})}$ are the corresponding multipole amplitudes, defined in the Appendix~\ref{app: multipole expansion}.

Upon substitution into Eq.~\eqref{eq: spectral density AMT}, angular integration over the closed surface can be carried out analytically. For instance, the contribution of the term in Eq.~\eqref{eq: cenital radial electric-field stress component} to the spectral momentum density is given by
\begin{align}
%
% Electric-Electric \varphi r sin \theta cos \varphi
%
\oint_S \mathcal{E}_{
\ell, m}^{(\text{e,s})\varphi}
\mathcal{E}_{\ell^{\prime}, m^{\prime}}^{(\text{e}^{\prime},\text{s}^{\prime})\text{r}\,*} \sin\theta\,d\Omega
= \, &\Bigg\{C_{\ell, m}^{\rm \text{(e,s)}}D_{\ell^{\prime}, m^{\prime}}^{\rm (\text{e}^{\prime},\text{s}^{\prime}) \, *} \frac{Z_{\ell}^{\rm (e,s)}Z_{\ell^{\prime}}^{\rm (\text{e}^{\prime},\text{s}^{\prime}) \, *}}{k r} \nonumber\\
&\times \Bigg[(\ell +1) IM_{\ell,\ell^{\prime}}^{m,m^{\prime}} \nonumber \\
& - (\ell -m + 1) \Delta_{\ell+1,\ell^{\prime}}^{m}\Bigg] \nonumber \\
& +D_{\ell, m}^{\rm \text{(e,s)}}D_{\ell^{\prime}, m^{\prime}}^{\rm (\text{e}^{\prime},\text{s}^{\prime}) \, *} m \nonumber \\
& \times \frac{f_{\ell}^{\rm (e,s)}Z_{\ell^{\prime}}^{\rm (\text{e}^{\prime},\text{s}^{\prime}) \, *}}{k r} \Delta_{\ell,\ell^{\prime}}^{m} \Bigg\} \nonumber \\
& \times i\, \ell^{\prime} (\ell^{\prime}+1)
2\pi\delta_{m,m^{\prime}}.
\end{align}

Here, the quantity
\begin{equation}
    IM_{\ell,\ell^{\prime}}^{m,m^{\prime}} = \int_{-1}^1 P_{\ell}^m (x) P_{\ell^{\prime}}^{m^{\prime}} (x) x\,dx,
\end{equation}
is one of the irreducible integrals introduced from now on, while
\begin{align}
    \Delta_{\ell,\ell^{\prime}}^{m} &= \int_{-1}^1 P_{\ell}^m (x) P_{\ell^{\prime}}^{m} (x)\, dx \nonumber \\
    &= \frac{2(\ell+m)!}{(2\ell+1)(\ell-m)!}\,\delta_{\ell,\ell^{\prime}}
\end{align}
is the associated-Legendre normalization integral, proportional to the Kronecker delta $\delta_{\ell,\ell^{\prime}}$ by the orthogonality of $P_\ell^m$ at fixed order $m$. Both $IM_{\ell,\ell^{\prime}}^{m,m^{\prime}}$ and $\Delta_{\ell,\ell^{\prime}}^{m}$ depend only on the integers $\ell,\ell^{\prime},m,m^{\prime}$ of the associated Legendre functions $P_{\ell}^m (x)$ and are independent of the parameters that describe the swift electron or the NP; $\Delta_{\ell,\ell^{\prime}}^{m}$ appears only in terms where azimuthal symmetry already forces $m^{\prime}=m$, as made explicit by the Kronecker delta $\delta_{m,m^{\prime}}$ multiplying every term above.

The remaining field-component products and their angular integrals follow by the same procedure and introduce three further irreducible integrals, $IU$, $IV$, and $IW$. We collect below the complete set of terms needed for the Cartesian $y$ component used throughout this work.

The electric-field contributions to the radial projection of the tensor in Eq.~\eqref{eq: spectral density AMT} can be written as follows. For the Cartesian $y$ component, one finds the following
\begin{align}
%
%   y component
%
\mathcal{L}^{\rm E}  = & \,\epsilon_0 \frac{R^3}{\pi}\sum_{\ell, m}\sum_{\ell^\prime,m^\prime} \int_0^{4\pi}\Re \Bigg\{ \mathcal{E}_{\ell, m}^{\text{(e,s)}\theta} \mathcal{E}_{\ell^{\prime}, m^{\prime}}^{(\text{e}^{\prime},\text{s}^{\prime})r*} \cos\varphi \nonumber \\
& -\mathcal{E}_{\ell, m}^{\text{(e,s)}\varphi} \mathcal{E}_{\ell^{\prime}, m^{\prime}}^{(\text{e}^{\prime},\text{s}^{\prime})r*} \cos\theta\sin\varphi \Bigg\} \,d\Omega.
\label{eq: projection Maxwell Tensor}
\end{align}
The $x$ and $z$ projections  can be found analogously. The magnetic-field contributions are obtained by replacing
$\epsilon_0 \rightarrow \mu_0$ and
$\mathcal{E}_{\ell m} \rightarrow \mathcal{H}_{\ell m}$ in Eq. \eqref{eq: projection Maxwell Tensor}.

The closed surface integral over the solid angle can be evaluated analytically in Eq. \eqref{eq: projection Maxwell Tensor}. The resulting expressions can be written as sums of angular integrals involving products of multipolar field components. Representative terms are listed below.
\begin{widetext}
\begin{align}
%
% Electric-Electric \theta r \sin\varphi
%
\oint_S\mathcal{E}_{\ell, m}^{\text{(e,s)}\theta}\mathcal{E}_{\ell^{\prime}, m^{\prime}}^{(\text{e}^{\prime},\text{s}^{\prime})r\,*} \sin\varphi \, d\Omega
= \, &\rmi \pi \left(\delta_{m-1, m^{\prime}} - \delta_{m+1, m^{\prime}}\right) \Bigg\{-C_{\ell,m}^{\rm (e,s)}D_{\ell^{\prime}, m^{\prime}}^{\rm (e^{\prime},s^{\prime}) \, *}\frac{Z_{\ell}^{\rm (e,s)}Z_{\ell^{\prime}}^{\rm (e^{\prime},s^{\prime}) \, *}}{k R} m \, IU_{\ell, \ell^{\prime}}^{m, m^{\prime}} \nonumber\\
&-D_{\ell, m}^{\rm (e,s)}D_{\ell^{\prime}, m^{\prime}}^{\rm (e^{\prime},s^{\prime}) \, *} \frac{f_{\ell}^{\rm (e,s)}Z_{\ell^{\prime}}^{\rm (e^{\prime},s^{\prime}) \, *}}{k R} \Bigg[ (\ell+1) IW_{\ell, \ell^{\prime}}^{m, m^{\prime}} \nonumber \\
&-(\ell-m+1) IU_{\ell + 1, \ell^{\prime}}^{m, m^{\prime}} \Bigg] \Bigg\}\,\ell^{\prime}(\ell^{\prime}+1),\\
%
%\end{align}
%
%\begin{align}
%
% Magnetic-Magnetic \theta r \sin\varphi
%
%\oint_S\mathcal{H}_{\ell, m}^{\text{(e,s)}\theta}\mathcal{H}_{\ell^{\prime}, m^{\prime}}^{(\text{e}^{\prime},\text{s}^{\prime})r\,*} \sin\varphi \, d\Omega
%= \, &\rmi \pi \left(\delta_{m-1, m^{\prime}} - \delta_{m+1, m^{\prime}}\right) \Bigg\{D_{\ell,m}^{\rm (e,s)}C_{\ell^{\prime}, m^{\prime}}^{\rm (e^{\prime},s^{\prime}) \, *}\frac{Z_{\ell}^{\rm (e,s)}Z_{\ell^{\prime}}^{\rm (e^{\prime},s^{\prime}) \, *}}{k R} m \, IU_{\ell, \ell^{\prime}}^{m, m^{\prime}} \nonumber\\
%
%&-C_{\ell, m}^{\rm (e,s)}C_{\ell^{\prime}, m^{\prime}}^{\rm (e^{\prime},s^{\prime}) \, *} \frac{f_{\ell}^{\rm (e,s)}Z_{\ell^{\prime}}^{\rm (e^{\prime},s^{\prime}) \, *}}{k R} \Bigg[ (\ell+1) IW_{\ell, \ell^{\prime}}^{m, m^{\prime}} -(\ell-m+1) IU_{\ell + 1, \ell^{\prime}}^{m, m^{\prime}} \Bigg] \Bigg\}\,\ell^{\prime}(\ell^{\prime}+1),
%
%\end{align}
%
%\begin{align}
%
% Electric-Electric \theta r \cos\varphi
%
\oint_S\mathcal{E}_{\ell, m}^{\text{(e,s)}\theta}\mathcal{E}_{\ell^{\prime}, m^{\prime}}^{(\text{e}^{\prime},\text{s}^{\prime})r\,*} \cos\varphi \, d\Omega
= \, &\pi \left(\delta_{m-1, m^{\prime}} + \delta_{m+1, m^{\prime}}\right) \Bigg\{-C_{\ell,m}^{\rm (e,s)}D_{\ell^{\prime}, m^{\prime}}^{\rm (e^{\prime},s^{\prime}) \, *}\frac{Z_{\ell}^{\rm (e,s)}Z_{\ell^{\prime}}^{\rm (e^{\prime},s^{\prime}) \, *}}{k R} m \, IU_{\ell, \ell^{\prime}}^{m, m^{\prime}} \nonumber\\
&-D_{\ell, m}^{\rm (e,s)}D_{\ell^{\prime}, m^{\prime}}^{\rm (e^{\prime},s^{\prime}) \, *} \frac{f_{\ell}^{\rm (e,s)}Z_{\ell^{\prime}}^{\rm (e^{\prime},s^{\prime}) \, *}}{k R} \Bigg[ (\ell+1) IW_{\ell, \ell^{\prime}}^{m, m^{\prime}} \nonumber \\
&-(\ell-m+1) IU_{\ell + 1, \ell^{\prime}}^{m, m^{\prime}} \Bigg] \Bigg\}\,\ell^{\prime}(\ell^{\prime}+1),
\end{align}
%
%\begin{align}
%
% Magnetic-Magnetic \theta r \cos\varphi
%
%\oint_S\mathcal{H}_{\ell, m}^{\text{(e,s)}\theta}\mathcal{H}_{\ell^{\prime}, m^{\prime}}^{(\text{e}^{\prime},\text{s}^{\prime})r\,*} \cos\varphi \, d\Omega
%= \, &\pi \left(\delta_{m-1, m^{\prime}} + \delta_{m+1, m^{\prime}}\right) \Bigg\{D_{\ell,m}^{\rm (e,s)}C_{\ell^{\prime}, m^{\prime}}^{\rm (e^{\prime},s^{\prime}) \, *}\frac{Z_{\ell}^{\rm (e,s)}Z_{\ell^{\prime}}^{\rm (e^{\prime},s^{\prime}) \, *}}{k R} m \, IU_{\ell, \ell^{\prime}}^{m, m^{\prime}} \nonumber\\
%
%&-C_{\ell, m}^{\rm (e,s)}C_{\ell^{\prime}, m^{\prime}}^{\rm (e^{\prime},s^{\prime}) \, *} \frac{f_{\ell}^{\rm (e,s)}Z_{\ell^{\prime}}^{\rm (e^{\prime},s^{\prime}) \, *}}{k R} \Bigg[ (\ell+1) IW_{\ell, \ell^{\prime}}^{m, m^{\prime}} -(\ell-m+1) IU_{\ell + 1, \ell^{\prime}}^{m, m^{\prime}} \Bigg] \Bigg\}\,\ell^{\prime}(\ell^{\prime}+1),
%
%\end{align}
%
\begin{align}
%
% Electric-Electric \varphi r \cos\theta\sin\varphi
%
\oint_S\mathcal{E}_{\ell, m}^{\text{(e,s)}\varphi}\mathcal{E}_{\ell^{\prime}, m^{\prime}}^{(\text{e}^{\prime},\text{s}^{\prime})r\,*}\cos\theta\sin\varphi \, d\Omega
= \, &-\pi \left(\delta_{m-1, m^{\prime}} - \delta_{m+1, m^{\prime}}\right) \Bigg\{ C_{\ell, m}^{\rm (e,s)} D_{\ell^{\prime}, m^{\prime}}^{\rm (e^{\prime},s^{\prime}) \, *} \frac{Z_{\ell}^{\rm (e,s)} Z_{\ell^{\prime}}^{\rm (e^{\prime},s^{\prime}) \, *}}{k R} \Bigg[(\ell +1) IV_{\ell, \ell^{\prime}}^{m, m^{\prime}} \nonumber \\
&-(\ell - m +1) IW_{\ell + 1, \ell^{\prime}}^{m, m^{\prime}} \Bigg] \nonumber\\
&+ D_{\ell, m}^{\rm (e,s)} D_{\ell^{\prime}, m^{\prime}}^{\rm (e^{\prime},s^{\prime}) \, *}\, m \frac{f_{\ell}^{\rm (e,s)} Z_{\ell^{\prime}}^{\rm (e^{\prime},s^{\prime})\, *}}{k R} IW_{\ell, \ell^{\prime}}^{m, m^{\prime}} \Bigg\} \, \ell^{\prime}(\ell^{\prime}+1),
\end{align}
%
%\begin{align}
%
% Magnetic-Magnetic \varphi r \cos\theta\sin\varphi
%
%\oint_S\mathcal{H}_{\ell, m}^{\text{(e,s)}\varphi}\mathcal{H}_{\ell^{\prime}, m^{\prime}}^{(\text{e}^{\prime},\text{s}^{\prime})r\,*}\cos\theta\sin\varphi \, d\Omega
%= \, &-\pi \left(\delta_{m-1, m^{\prime}} - \delta_{m+1, m^{\prime}}\right) \Bigg\{ -D_{\ell, m}^{\rm (e,s)} C_{\ell^{\prime}, m^{\prime}}^{\rm (e^{\prime},s^{\prime}) \, *} \frac{Z_{\ell}^{\rm (e,s)} Z_{\ell^{\prime}}^{\rm (e^{\prime},s^{\prime}) \, *}}{k R} \Bigg[(\ell +1) IV_{\ell, \ell^{\prime}}^{m, m^{\prime}} -(\ell - m +1) IW_{\ell + 1, \ell^{\prime}}^{m, m^{\prime}} \Bigg] \nonumber\\
%
%&+ C_{\ell, m}^{\rm (e,s)} C_{\ell^{\prime}, m^{\prime}}^{\rm (e^{\prime},s^{\prime}) \, *}\, m \frac{f_{\ell}^{\rm (e,s)} Z_{\ell^{\prime}}^{\rm (e^{\prime},s^{\prime})\, *}}{k R} IW_{\ell, \ell^{\prime}}^{m, m^{\prime}} \Bigg\} \, \ell^{\prime}(\ell^{\prime}+1),
%
%\end{align}
%
\begin{align}
%
% Electric-Electric \varphi r \cos\theta\cos\varphi
%
\oint_S\mathcal{E}_{\ell, m}^{\text{(e,s)}\varphi}\mathcal{E}_{\ell^{\prime}, m^{\prime}}^{(\text{e}^{\prime},\text{s}^{\prime})r\,*}\cos\theta\cos\varphi \, d\Omega
= \, &\rmi \, \pi \left(\delta_{m-1, m^{\prime}} + \delta_{m+1, m^{\prime}}\right) \Bigg\{ C_{\ell, m}^{\rm (e,s)} D_{\ell^{\prime}, m^{\prime}}^{\rm (e^{\prime},s^{\prime}) \, *} \frac{Z_{\ell}^{\rm (e,s)} Z_{\ell^{\prime}}^{\rm (e^{\prime},s^{\prime}) \, *}}{k R} \Bigg[(\ell +1) IV_{\ell, \ell^{\prime}}^{m, m^{\prime}} \nonumber \\
&-(\ell - m +1) IW_{\ell + 1, \ell^{\prime}}^{m, m^{\prime}} \Bigg] \nonumber\\
&+ D_{\ell, m}^{\rm (e,s)} D_{\ell^{\prime}, m^{\prime}}^{\rm (e^{\prime},s^{\prime}) \, *}\, m \frac{f_{\ell}^{\rm (e,s)} Z_{\ell^{\prime}}^{\rm (e^{\prime},s^{\prime})\, *}}{k R} IW_{\ell, \ell^{\prime}}^{m, m^{\prime}} \Bigg\}\, \ell^{\prime}(\ell^{\prime}+1),
\end{align}
%
%\begin{align}
%
% Magnetic-Magnetic \varphi r \cos\theta\cos\varphi
%
%\oint_S\mathcal{H}_{\ell, m}^{\text{(e,s)}\varphi}\mathcal{H}_{\ell^{\prime}, m^{\prime}}^{(\text{e}^{\prime},\text{s}^{\prime})r\,*}\cos\theta\cos\varphi \, d\Omega
%= \, &\rmi \, \pi \left(\delta_{m-1, m^{\prime}} + \delta_{m+1, m^{\prime}}\right) \Bigg\{ -D_{\ell, m}^{\rm (e,s)} C_{\ell^{\prime}, m^{\prime}}^{\rm (e^{\prime},s^{\prime}) \, *} \frac{Z_{\ell}^{\rm (e,s)} Z_{\ell^{\prime}}^{\rm (e^{\prime},s^{\prime}) \, *}}{k R} \Bigg[(\ell +1) IV_{\ell, \ell^{\prime}}^{m, m^{\prime}} -(\ell - m +1) IW_{\ell + 1, \ell^{\prime}}^{m, m^{\prime}} \Bigg] \nonumber\\
%
%&+ C_{\ell, m}^{\rm (e,s)} C_{\ell^{\prime}, m^{\prime}}^{\rm (e^{\prime},s^{\prime}) \, *}\, m \frac{f_{\ell}^{\rm (e,s)} Z_{\ell^{\prime}}^{\rm (e^{\prime},s^{\prime})\, *}}{k R} IW_{\ell, \ell^{\prime}}^{m, m^{\prime}} \Bigg\}\, \ell^{\prime}(\ell^{\prime}+1),
%
%\end{align}
%
\begin{align}
%
% Electric-Electric \varphi r \sin\theta
%
\oint_S\mathcal{E}_{\ell, m}^{\text{(e,s)}\varphi}\mathcal{E}_{\ell^{\prime}, m^{\prime}}^{(\text{e}^{\prime},\text{s}^{\prime})r\,*}\sin\theta \, d\Omega
= \, &\rmi \, 2\pi \, \delta_{m m^{\prime}} \Bigg\{ C_{\ell, m}^{\rm (e,s)} D_{\ell^{\prime}, m^{\prime}}^{\rm (e^{\prime},s^{\prime}) \, *} \frac{Z_{\ell}^{\rm (e,s)} Z_{\ell^{\prime}}^{\rm (e^{\prime},s^{\prime}) \, *}}{k R} \Bigg[(\ell +1) IM_{\ell, \ell^{\prime}}^{m, m^{\prime}} -(\ell - m +1) \Delta_{\ell + 1, \ell^{\prime}} \Bigg] \nonumber\\
&+ D_{\ell, m}^{\rm (e,s)} D_{\ell^{\prime}, m^{\prime}}^{\rm (e^{\prime},s^{\prime}) \, *}\, m \frac{f_{\ell}^{\rm (e,s)} Z_{\ell^{\prime}}^{\rm (e^{\prime},s^{\prime})\, *}}{k R} \Delta_{\ell \ell^{\prime}} \Bigg\} \, \ell^{\prime}(\ell^{\prime}+1).
\end{align}
%
%\begin{align}
%
% Magnetic-Magnetic \varphi r \sin\theta
%
%\oint_S\mathcal{H}_{\ell, m}^{\text{(e,s)}\varphi}\mathcal{H}_{\ell^{\prime}, m^{\prime}}^{(\text{e}^{\prime},\text{s}^{\prime})r\,*}\sin\theta \, d\Omega
%= \, &\rmi \, 2\pi \, \delta_{m m^{\prime}} \Bigg\{ -D_{\ell, m}^{\rm (e,s)} C_{\ell^{\prime}, m^{\prime}}^{\rm (e^{\prime},s^{\prime}) \, *} \frac{Z_{\ell}^{\rm (e,s)} Z_{\ell^{\prime}}^{\rm (e^{\prime},s^{\prime}) \, *}}{k R} \Bigg[(\ell +1) IM_{\ell, \ell^{\prime}}^{m, m^{\prime}} -(\ell - m +1) \Delta_{\ell + 1, \ell^{\prime}} \Bigg] \nonumber\\
%
%&+ C_{\ell, m}^{\rm (e,s)} C_{\ell^{\prime}, m^{\prime}}^{\rm (e^{\prime},s^{\prime}) \, *}\, m \frac{f_{\ell}^{\rm (e,s)} Z_{\ell^{\prime}}^{\rm (e^{\prime},s^{\prime})\, *}}{k R} \Delta_{\ell \ell^{\prime}} \Bigg\} \, \ell^{\prime}(\ell^{\prime}+1).
%
%\end{align}
%
\end{widetext}
All remaining angular integrals can be derived by following the same procedure, and can be found in Ref. \cite{briseno2023masterthesis}. Together, these expressions provide a complete analytical evaluation of the angular-momentum-transfer spectral density.

The expressions presented here are included to illustrate the underlying methodology and to identify the set of irreducible integrals that appear in the full expression of the spectral density $\mathcal{L}$. The complete set of these irreducible integrals is:
\vspace{-0.07cm}
\begin{align}
    %IN_{\ell,\ell^{\prime}}^{m,m^{\prime}} =& \int_{-1}^1 P_{\ell}^m (x) P_{\ell^{\prime}}^{m^{\prime}} (x) \sqrt{1-x^2}\,dx,
    %\\
    IM_{\ell,\ell^{\prime}}^{m,m^{\prime}} =& \int_{-1}^1 P_{\ell}^m (x) P_{\ell^{\prime}}^{m^{\prime}} (x) \,x\,dx,
    \\
    IU_{\ell,\ell^{\prime}}^{m,m^{\prime}} =& \int_{-1}^1 P_{\ell}^m (x) P_{\ell^{\prime}}^{m^{\prime}} (x) \frac{1}{\sqrt{1-x^2}}\,dx,
    \\
    IV_{\ell,\ell^{\prime}}^{m,m^{\prime}} =& \int_{-1}^1 P_{\ell}^m (x) P_{\ell^{\prime}}^{m^{\prime}} (x) \frac{x^2}{\sqrt{1-x^2}}\,dx,
    \\
    IW_{\ell,\ell^{\prime}}^{m,m^{\prime}} =& \int_{-1}^1 P_{\ell}^m (x) P_{\ell^{\prime}}^{m^{\prime}} (x) \frac{x}{\sqrt{1-x^2}}\,dx,
    %\\
    %IX_{\ell,\ell^{\prime}}^{m,m^{\prime}} =& \int_{-1}^1 P_{\ell}^m (x) P_{\ell^{\prime}}^{m^{\prime}} (x) \frac{x^2}{1-x^2}\,dx,
    %\\
    %IY_{\ell,\ell^{\prime}}^{m,m^{\prime}} =& \int_{-1}^1 P_{\ell}^m (x) P_{\ell^{\prime}}^{m^{\prime}} (x) \frac{x}{1-x^2}\,dx, \\
    %IZ_{\ell,\ell^{\prime}}^{m,m^{\prime}} =& \int_{-1}^1 P_{\ell}^m (x) P_{\ell^{\prime}}^{m^{\prime}} (x) \frac{x^3}{1-x^2}\,dx,
    \\
    \Delta_{\ell,\ell^{\prime}}^m =& \int_{-1}^1 P_{\ell}^m (x) P_{\ell^{\prime}}^m(x) \, dx \nonumber \\
    =& \frac{2(\ell+m)!}{(2\ell+1)(\ell-m)!} \delta_{\ell,\ell^{\prime}},
\end{align}
which can be evaluated to machine precision using Gauss-Legendre and Gauss-Chebyshev quadratures \cite{kahaner1989numerical,press2007numerical}.

The integral of $\mathcal{L}$ on the frequency space is performed employing the Gauss-Kronrod quadrature, as described in Ref.~\cite{castrejon2021effects}.

%%%%%%%%%%%%%%%%%%%%%%%%%%%%%%%%%%%%%%%%%%%%%%%%%%%%%%%%%%%%%%%%%%%%%%%%%
\section{Dielectric function of gold}\label{app: dielectric Au}
%%%%%%%%%%%%%%%%%%%%%%%%%%%%%%%%%%%%%%%%%%%%%%%%%%%%%%%%%%%%%%%%%%%%%%%%%

The dielectric response of gold is modeled using a sum of one Drude term and eight Lorentz oscillators, yielding
\begin{equation}
    \epsilon(\omega) = 1 + \sum_{n=1}^{9}
    \frac{A_n}{\omega_{n}^2 - \omega^2 - i\omega\Gamma_n} \,,
\end{equation}
where $\omega_n$ and $\Gamma_n$ are the resonance frequency and damping coefficient of the $n$th oscillator, respectively ($\omega_1=0$ corresponds to the Drude term), and $A_n$ is the corresponding oscillator strength. All energies are expressed in electronvolts (eV), with $\hbar$ absorbed for brevity. This parametrization follows the same functional form used for bismuth and gold alike in Ref.~\cite{werner}, from which the fit reproducing gold's REELS-derived optical response is taken. As emphasized by Werner \emph{et al.}~\cite{werner} themselves, the individual oscillator energies in Table~\ref{tab:gold_params} are phenomenological fit parameters rather than direct spectroscopic assignments; nonetheless, the lowest-energy Lorentz term ($\hbar\omega_2=4.0$~eV) coincides closely with the interband peak reported in optical measurements~\cite{johnsonchristy1972}.
\begin{table}[h!]
\caption{\label{tab:gold_params}%
Lorentz-oscillator parameters used for the dielectric function of gold (Ref.~\cite{werner}); $n=1$ is the Drude term.
}
\begin{ruledtabular}
\begin{tabular}{cccc}
$n$ & $\hbar\omega_{n}$ (eV) & $\hbar\Gamma_{n}$ (eV) & $\hbar^2 A_{n}$ (eV$^{2}$) \\
\hline
1 & 0.0  & 0.2  & 113.1 \\
2 & 4.0  & 1.5  & 44.6  \\
3 & 7.3  & 3.3  & 54.8  \\
4 & 12.8 & 11.8 & 184.9 \\
5 & 18.9 & 71.0 & 728.1 \\
6 & 19.9 & 2.9  & 65.7  \\
7 & 28.9 & 3.9  & 50.0  \\
8 & 38.7 & 13.0 & 74.7  \\
9 & 64.3 & 51.9 & 544.0 \\
\end{tabular}
\end{ruledtabular}
\end{table}

%%%%%%%%%%%%%%%%%%%%%%%%%%%%%%%%%%%%%%%%%%%%%%%%%%%%%%%%%%%%%
\section{Numerical implementation}\label{app:numerical}
%%%%%%%%%%%%%%%%%%%%%%%%%%%%%%%%%%%%%%%%%%%%%%%%%%%%%%%%%%%%%%

The numerical implementation of the expressions developed in this work is openly available on GitHub~\cite{AMTRepo}.
To reproduce Figs.~\ref{fig:spectral_density}-\ref{fig:Al_vs_Au}, the repository must be built locally, after which the simulations can be executed using, e.g.,
\begin{verbatim}
./DrudeAl_AMTsolver -a 30 --vscan
\end{verbatim}
or
\begin{verbatim}
./DrudeAl_AMTsolver -a 30 --bscan
\end{verbatim}
with the nanoparticle radius \texttt{-a} set to the value quoted in the corresponding figure caption ($5$, $20$, $30$, or $50$~nm), and analogously with the gold dielectric function to reproduce Figs.~\ref{fig:maps_Au30}a), \ref{fig:maps_Au30}b), \ref{fig:DL_v_Au}, and~\ref{fig:Al_vs_Au}b).

Each execution generates an output directory organized by material, nanoparticle radius, and execution timestamp.
For a given run, the results are separated into folders corresponding to speed scans at fixed impact parameter and impact parameter scans at fixed electron speed.
The speed-scan results are further grouped according to the maximum multipole order, \(\ell_{\max}\), and include the computed angular momentum transfer together with the associated numerical error estimates.
Additional files record the multipolar convergence analysis.
Each simulation directory also contains a text file summarizing the input parameters and metadata associated with the run.

Further details on compilation, runtime parameters available, and input options are provided in the repository \texttt{README} file.

%
%
% REFERENCES
%
%

\bibliographystyle{apsrev4-2}
\bibliography{references}

\end{document}